\documentclass[a4paper]{spie}
\usepackage[]{graphicx}
\title{Phase Referencing in Optical Interferometry}

\author{Mercedes E. Filho\supit{a}, Paulo Garcia\supit{a,b}, Gilles Duvert\supit{c}, Gaspard Duchene\supit{d}, Eric Thiebaut\supit{e}, John Young\supit{f}, Olivier Absil\supit{g},
Jean-Phillipe Berger\supit{g},
Thomas Beckert\supit{h},
Sebastian Hoenig\supit{h},
Dieter Schertl\supit{h},
Gerd Weigelt\supit{h},
Leonardo Testi\supit{i},
Eric Tatuli\supit{i},
Virginie Borkowski\supit{j},
Micha\"el de Becker\supit{j},
Jean Surdej\supit{j},
Bernard Aringer\supit{k},
Joseph Hron\supit{k},
Thomas Lebzelter\supit{k},
Andrea  Chiavassa\supit{l},
Romano Corradi\supit{l},
Tim Harries\supit{m}
\skiplinehalf
\supit{a}Centro de Astrofisica da Universidade do Porto, Rua das Estrelas, 4150-762 Porto, Portugal \\
\supit{b} Departamento de Engenharia F\'{\i}sica, Faculdade de
Engenharia, Universidade do Porto, Portugal \\
\supit{c}Laboratoire d'Astrophysique de Grenoble, Observatoire de Grenoble, BP 53, 38041 Grenoble Cedex 9, France \\
\supit{d}UC Berkeley, Astronomy Department, 601 Campbell Hall, Berkeley CA 94720-3411, USA \\
\supit{e}Observatoire de Lyon, 9 Av. Charles Andr\'e, 69561 Saint Genis Laval Cedex, France \\
\supit{f}Cavendish Laboratory, Madingley Road, Cambridge CB3 OHE, UK\\
\supit{g}Universit\'e J. Fourier, CNRS, Laboratoire d'Astrophysique de Grenoble, UMR 5571, France;\\
\supit{h}Max-Planck Institute for Radioastronomy, Bonn, Germany;\\
\supit{i}INAF/Osservatorio di Astrofisica di Arcetri, Italy;\\
\supit{j}Institute of Astrophysics and Geophysics, Li\`ege, Belgium;\\
\supit{k}Institute of Astrophysics of the University of Wien, Austria;\\
\supit{l}Groupe  de  Recherche  en Astronomie  et Astrophysique du Languedoc,
Montpellier, France;\\
\supit{m}School of Physics, University of Exeter, UK;\\
}

\authorinfo{Further author information: (Send correspondence to M. E. F.)\\M. E. F.: E-mail: mfilho@astro.up.pt, Telephone: +351 6089 853}

\begin{document}
 \maketitle


\begin{abstract}

One of the aims of next generation optical interferometric instrumentation
is to be able to make use of information contained in the visibility phase to construct high dynamic range images.

Radio and optical interferometry are at the two extremes of phase corruption by the atmosphere.
While in radio it is possible to obtain calibrated phases for the science objects, in the optical this is currently not possible.
Instead, optical interferometry has relied on
closure phase techniques to produce images. Such techniques allow only to
achieve modest dynamic ranges.
However, with high contrast objects, for faint targets or when structure detail 
is needed, phase referencing techniques as used in radio interferometry, should theoretically achieve higher dynamic ranges for the same number of telescopes.

Our approach is not to provide evidence either for or against the hypothesis that phase referenced imaging gives better dynamic range than closure phase imaging. Instead we wish to explore the potential of this technique for future optical interferometry and also because image reconstruction in the optical using phase referencing techniques has only been performed with limited success. 

We have generated simulated, noisy, complex visibility data, analogous to the signal produced in radio interferometers, using the VLTI as a template. We proceeded with image reconstruction using the radio image reconstruction algorithms contained in {\sc aips} {\sc imagr} ({\sc clean} algorithm). Our results show that image reconstruction is successful in most of our science cases, yielding images with a 4 milliarcsecond resolution in K band. 

We have also investigated the number of target candidates for optical phase referencing. Using the 2MASS point source catalog, we show that there are several hundred objects with phase reference sources less than 30 arcseconds away, allowing to apply this technique.

\end{abstract}



\section{INTRODUCTION}
\label{sec:intro}  

An ideal interferometer will measure the complex visibility of an astronomical object. Image reconstruction deals with
 inverting the visibility information into an image, given poor Fourier plane sampling and limited phase information.

In an array of {\it N} telescopes, signals are combined in {\it $\frac{1}{2}$ $\times$ N $\times$ (N-1)} pairs or baselines to obtain {\it $\frac{1}{2}$ $\times$ N $\times$ (N-1)} measurements called complex visibilities. These visibilities are related to the object brightness distribution via the van Cittert-Zernike theorem:

\begin{center}
$V(u,v) = \int \int I(x,y) \, exp[-2\,  \pi \, i \, (ux+vy)] dx \, dy$
\end{center}

\noindent where {\it x} and {\it y} are angular displacements on the plane of the sky with the phase center as origin, {\it I(x,y)} is the brightness distribution of the target and {\it u} and {\it v} are the position vectors of the baselines projected on a plane perpendicular to the source direction, which together define the {\it uv} plane. In practical terms, the better the sampling of the {\it uv} plane in terms of baseline length, position angle, and number of measurements, the more faithful the reconstructed image will be relative to the true brightness distribution.

Radio-type observations are performed by measuring the amplitudes (modulus) and the phases (argument) of the complex visibilities:

\begin{center}
$V(u,v) = A \, exp (i\, \phi) $
\end{center}

\noindent Image reconstruction using the modulus and argument of the visibility function is called phase referencing image reconstruction and is widely used in radio astronomy. Studies of this type were previously attempted by Masoni (2006), Masoni et al. (2005) and Weigelt et al. (2008; private communication).

\section {Array Configuration}

Telescope configurations are an essential part of the image reconstruction process. We began by identifying array configurations that allowed the most uniform uv coverage:

\begin{figure}[ht!]
   \begin{center}
   \begin{tabular}{c}
  \includegraphics[height=7cm,angle=-90]{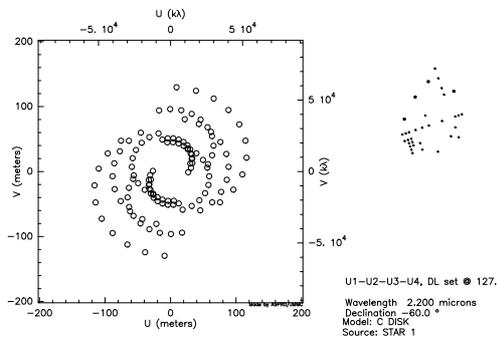}
   \end{tabular}
   \end{center}
   \caption[example]
   { \label{fig:example} 4 UT $\times$ 1 night configuration.}
\end{figure}

\begin{figure}[h!]
   \begin{center}
   \begin{tabular}{c}
   \includegraphics[height=7cm,angle=-90]{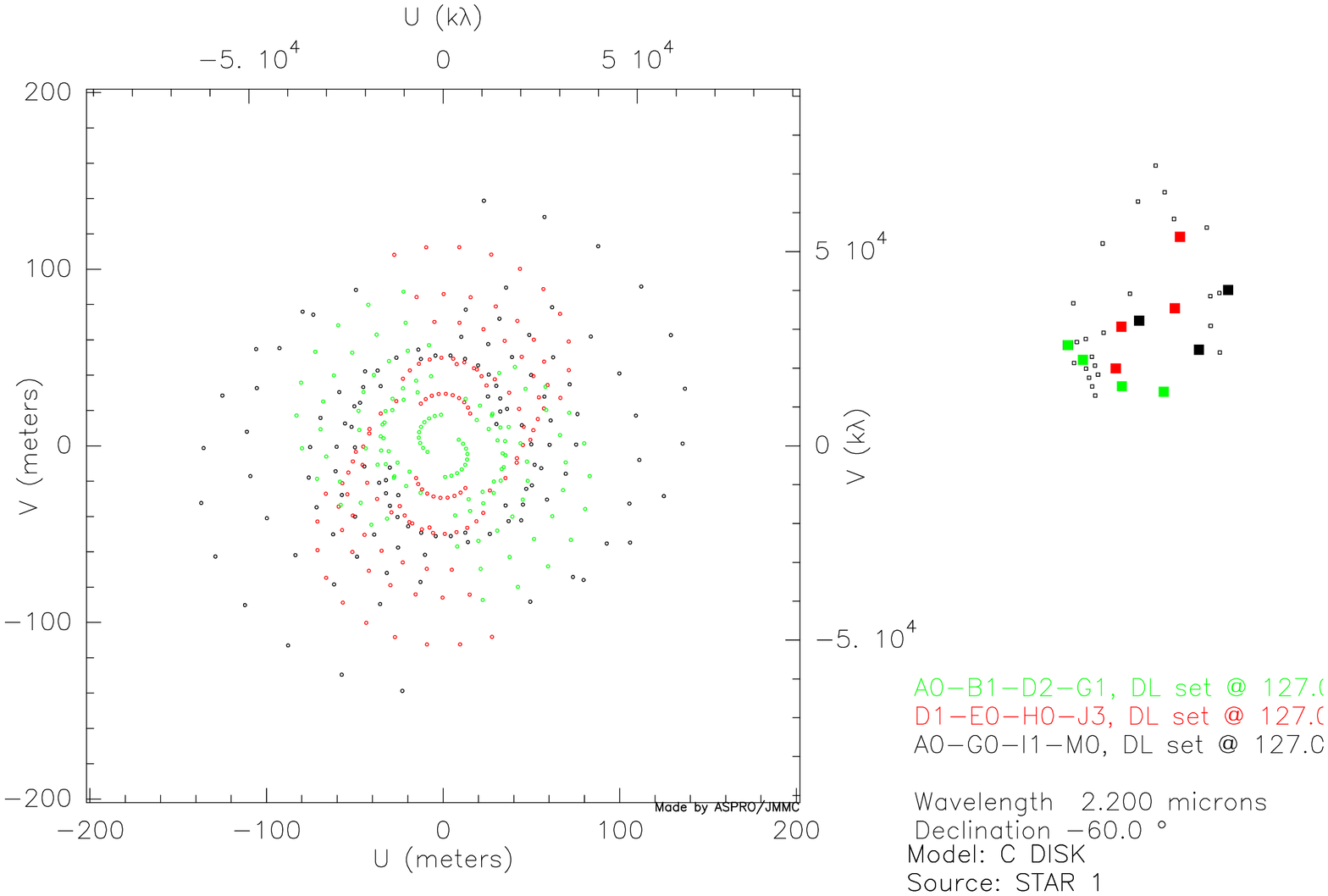}
   \includegraphics[height=7.5cm,angle=-90]{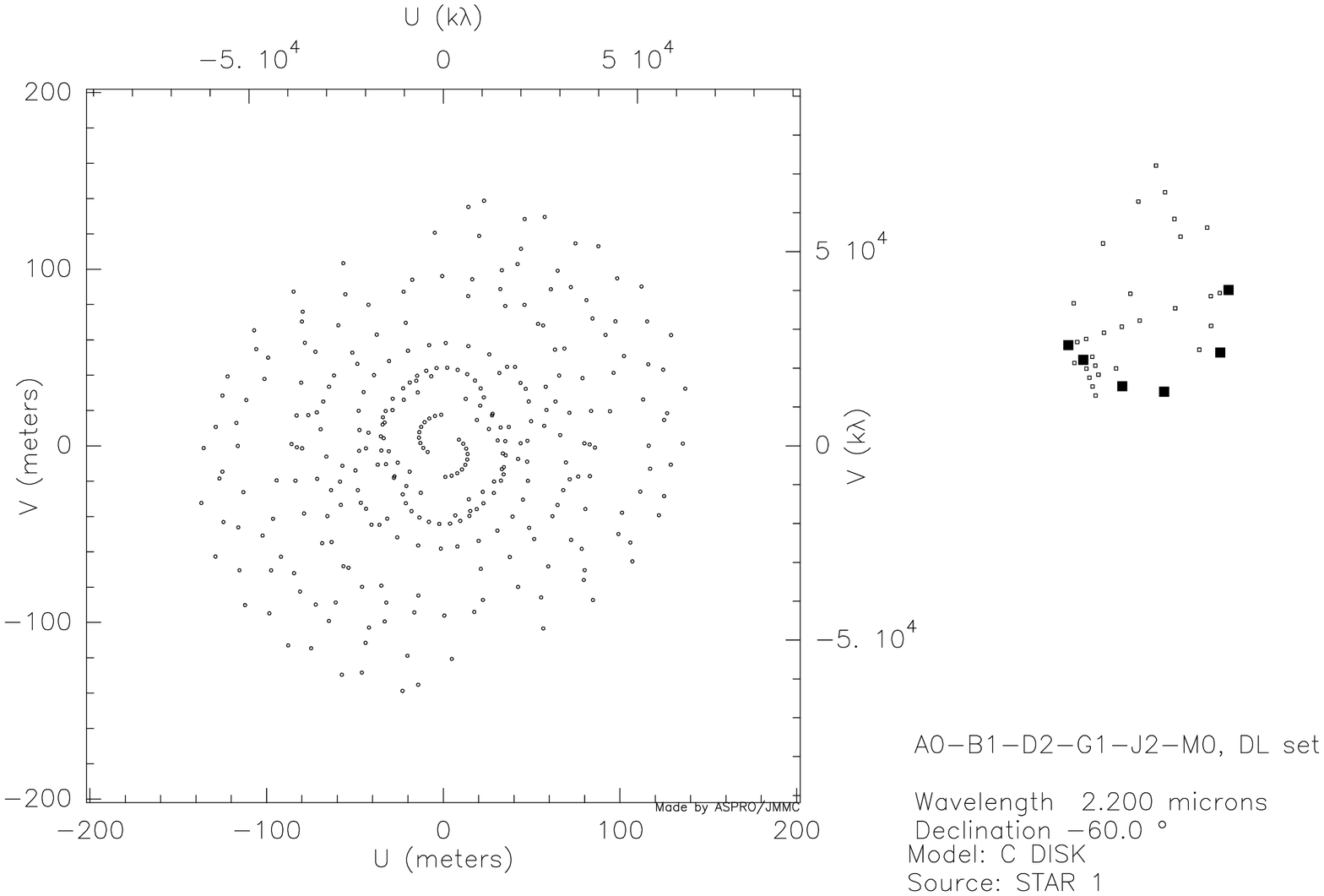}
   \end{tabular}
   \end{center}
   \caption[example]
   { \label{fig:example} 4 AT $\times$ 3 night configuration (left). 6 AT $\times$ 1 night configuration (right).}
\end{figure}

\begin{itemize}

\item 4 UTs $\times$ 1 night –- the chosen configuration when observing very faint sources (Fig. 1);

\item 4 ATs $\times$ 3 nights –- the case where there is a small number of telescopes (Fig. 2);

\item 6 ATs $\times$ 1 night –- a 6 telescope extended configuration with one night observation; has less $uv$ points than the 4 AT x 3 nights configuration but comparable $uv$ coverage (Fig. 3).

\end{itemize}

\section{Noise Model}

The noise model estimates the  uncertainties on the visibility amplitudes and phases assuming an instrument following a multi-axial recombination scheme with a fringe tracker (FT; Jocou et al. 2007).


The quantity we procure is the total number of detected photoevents per integration time per pixel per baseline in the interferometric channel:

\begin{center}
${\overline N_i} = \frac {N_{total} * f }{N_{pix} * N_{base}}$
\end{center}

\noindent where $f$ is the fraction of the beam that gores into the interferometric channel (90\%), $N_{pix}$ is the number of pixels needed to read the interferometric channel (600), $N_{base}$ is the number of baselines and $N_{total}$ is the total number of detected photevents per integration time per pixel in all channels:

\begin{center}
$N_{total} = F_0 * 10^{-0.4 mag} * t_{int} * N_{tel} * \pi * R^2 * \Delta \lambda * trans * Strehl$
\end{center}

\noindent Here $F_0$ is the photon flux of a zeroth magnitude star (Jocou et al. 2007), $mag$ is the object magnitude in the observing band, $t_{int}$ is the integration time, $N_{tel}$ is the number of telescopes (6 or 4), $R$ is the radius of the telescopes (4.1 for UTs and 0.9 for ATs) assumed for simplification to have no central hole, $\Delta \lambda$ is the spectral bandwidth (chosen), $trans$ is the total instrument transmission including quantum efficiency (Jocou et al. 2007), and $Strehl$ is the Strehl ratio and depends on wavelength (Jocou et al. 2007).

\noindent Therefore, the total number of detected photevents in the interferometric channel is:

\begin{center}
$N_i = {\overline N_i} * N_{pix} * N_{int}$
\end{center}

\noindent where $N_{int}$ is the number of independent integrations (depends on magnitude and FT presence).

The object intrinsic visibility, $V$, must be corrected for the instrumental visibility loss (80\%) and the instrumental visibility loss induced by the FT (90\%). The correlated flux per baseline is therefore:

\begin{center}
$F_{cor} = N_i *\frac {V'}{2}$
\end{center}

\noindent and finally the error in visibility is given by:

\begin{center}
$error_V= \sqrt{F_{cor} + N_{int} * N_{pix} * \sigma^2}$
\end{center}

\noindent where $\sigma$ os the readout noise of the detector (15 e-).

\section {UVFITS File Generation}

UVFITS is the file format in which radio astronomical data are written and used for phase referencing image reconstruction. UVFITS format is designed so that different categories of information are stored in distinct “tables” within a file and can be cross-referenced one to another. Each UVFITS file “table” stores specific parameters that include important interferometric observables and system information.

Key science images were generated and provided by the science case groups (Garcia et al. 2007).
Using {\sc aspro}, an image simulation tool originally created for IRAM, and the configurations above, K band "images" of the sources were created assuming that all sources were observed at the fixed declination of $-$60 degrees.

It was assumed also that during the night, the telescope configurations would remain fixed and that one calibrated {\it uv} point per baseline should be obtained every hour. The actual on-source integration time is, however, 10-15 minutes per hour due to overheads. The total integration time assumes an entire transit (9 hours).

\section {Phase Referencing Theory}

A radio interferometer works by phase referencing. The interferometer observes a science target and records the
visibility modulus and phase for each baseline. It also observes a reference target used to calibrate the visibility
modulus and phase for atmospheric variations. Therefore, as opposed to conventional optical interferometry, phase referencing makes
use of crucial information contained in the phases to recover the brightness distribution of a source.


\begin{figure}[h!]
   \begin{center}
   \begin{tabular}{c}
  \includegraphics[height=4cm]{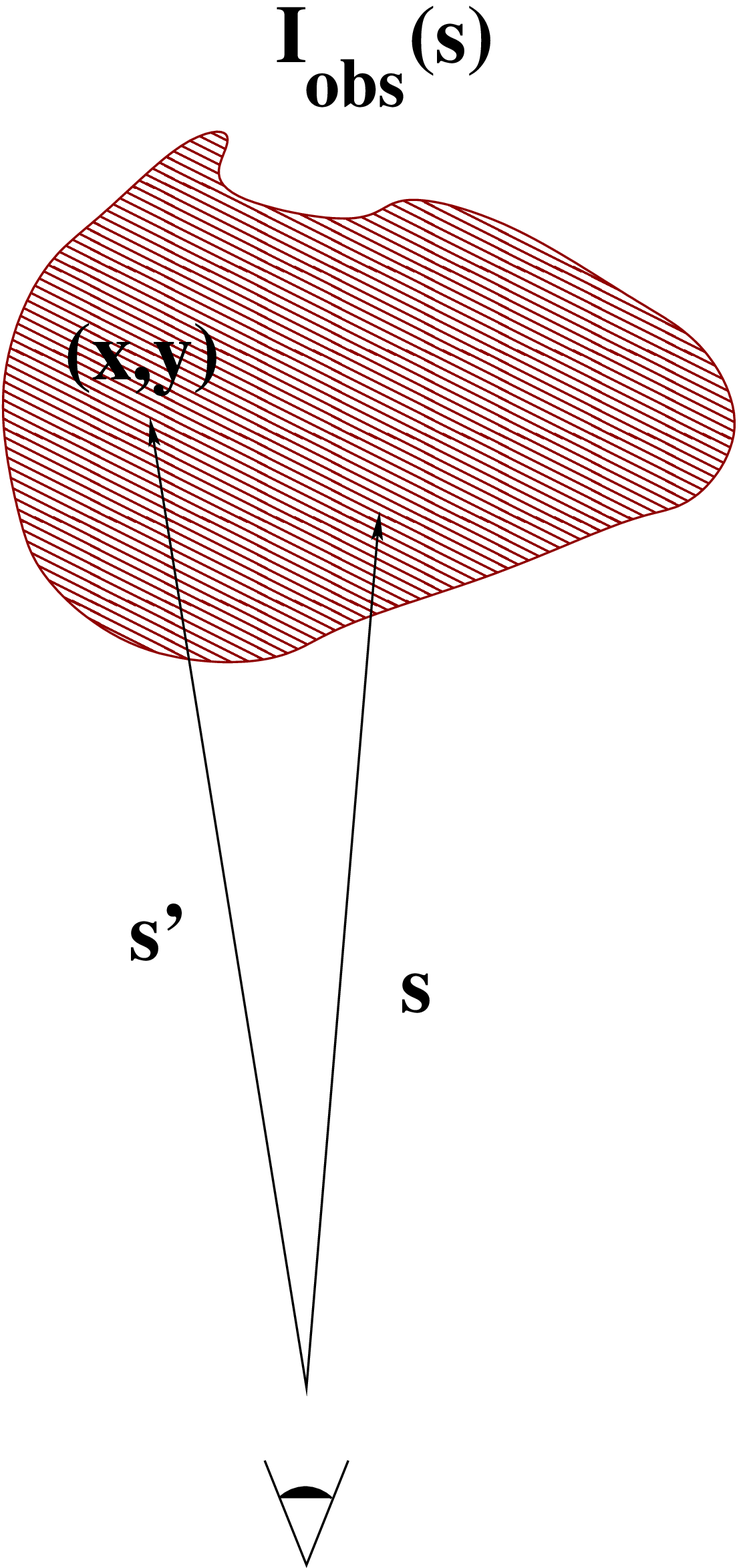}
   \end{tabular}
   \end{center}
   \caption[example]
   { \label{fig:example} }
   \end{figure}

Given an incoherent source, the observed brightness distribution in a direction $\vec {s'}$ is given by:

\begin{center}
$I^{obs}(\vec{s'}) = I^{obs}(x,y) = PSF(x,y) * I^{true}(x,y) + N(x,y)$
\end{center}

\noindent where {\it PSF(x,y)} is the instrumental point spread function, {\it I$^{true}$(x,y)} is the true object brightness distribution, {\it N(x,y)} is the noise and the asterisk denotes convolution.

In practice, interferometry does not make measurements in the image plane but in Fourier space. The relevant quantity is called the complex visibility and is measured at each {\it uv} point, the position vector of the baseline on a plane perpendicular to the source direction:

\begin{center}
$V^{obs}(u,v) = S(u,v) \times V^{true}(u,v) + N'(u,v)$
\end{center}

\noindent where {\it S(u,v)}, the sampling function, is the Fourier transform of the {\it PSF(x,y)}, {\it V$^{true}$(x,y)}, the true visibility, is the Fourier transform of the true brightness distribution {\it I$^{true}$(x,y)} and {\it N'(u,v)} is the noise in the Fourier space.

The van Cittert-Zernike theorem states that the true brightness distribution can be obtained by the inverse Fourier transform and deconvolution of the observables:

\begin{center}
$I^{true}(x,y) * PSF(x,y) = \int \int V^{true}(u,v) \times S(u,v) \, exp \, [2 \pi i (ux+vy)]\, du \, dv$
\end{center}

The role of image reconstruction is to obtain the best approximation, {\it I$^{aprox}$(x,y) $\sim$ I$^{obs}$(x,y) $\sim$ I$^{true}$(x,y)}, to the true brightness distribution.

In radio-like phase referencing image reconstruction, the data are gridded, interpolated and inverse Fourier transformed to yield a model representation of the sky:

\begin {center}
$I^{dirty}(x,y) = I^{aprox}(x,y) * PSF^{dirty}(x,y) + N(x,y)$
\end{center}

\noindent where $I^{dirty}(x,y)$ is called the dirty map and $PSF^{dirty}(x,y)$ is called the dirty beam.

The NRAO Astronomical Image Processing System ({\sc aips}), is a
baseline-based reconstruction method used in radio interferometry. The UVFILES generated by {\sc aspro} were imported into {\sc aips}. We have tested the results using the {\sc clean} algorithm (H\"{o}gbom 1974), which corrects for the effect of poor Fourier plane sampling. {\sc clean} grids, Fourier transforms and deconvolves the dirty beam from the dirty image in an iterative fashion given an initial guess for the beam and the brightness distribution. {\sc clean} then finds peaks in the residual image and subtracts $\delta$ functions of the appropriate strength at those positions. The final map is a convolution of all the $\delta$ functions with a {\sc clean} beam plus the residual map.

\section {Image Analysis}

In order to compare the reconstructed with the synthetic images, {\sc aspro} was
used to generate the point spread functions (PSF) for the 6 AT $\times$ 1 night, 4 AT $\times$ 3 nights and 4 UT $\times$ 1 night configurations (Table 1). The synthetic images were then convolved with a Gaussian of the measured PSF parameters ({\sc iraf} program {\sc gauss}). 

\begin{table}[h!]
\begin{center}
\begin{minipage}[c]{86mm}
\footnotesize
\caption{PSF parameters.
Col. 1: Configuration.
Col. 2: Full width at half maximum.
Col. 3: Dispersion, $\sigma=FWHM/2.35$.
Col. 4: Elllipticity, $e=\frac{a-b}{a+b}$, where $a$ is the major and $b$ the minor axis.
Col. 5: Axis ratio, $r=\frac{1-e}{1+e}$.
Col. 6: Position angle of the minor axis.}

\begin{tabular} {l | c c c c c }

\hline
Configuration   & FWHM  & $\sigma$ & e & r & PA \\

\hline

4 UT $\times$ 1 night & 4.45 & 1.89 & 0.08 & 0.85 & $-$9.0$^{\cdot}$ \\

4 AT $\times$ 3 nights & 3.51 & 1.49 & 0.38 & 0.45 & $-$83.0$^{\cdot}$ \\

6 AT $\times$ 1 night & 5.17 & 2.20 & 0.42 & 0.41 & $-$85.0$^{\cdot}$ \\

\hline

\end{tabular}
\end{minipage}
\end{center}
\end{table}

Relative astrometry information was obtained for the images using {\sc ds9}. Photometry of the image components was performed using {\sc iraf} procedure {\sc phot}. $SNR$ is the ratio of the mean pixel value to the standard deviation meansured on the reconstructed images.










\section{Phase Referencing Image Reconstruction}

\begin{figure}[h!]
   \begin{center}
   \begin{tabular}{c}
   \includegraphics[height=4cm]{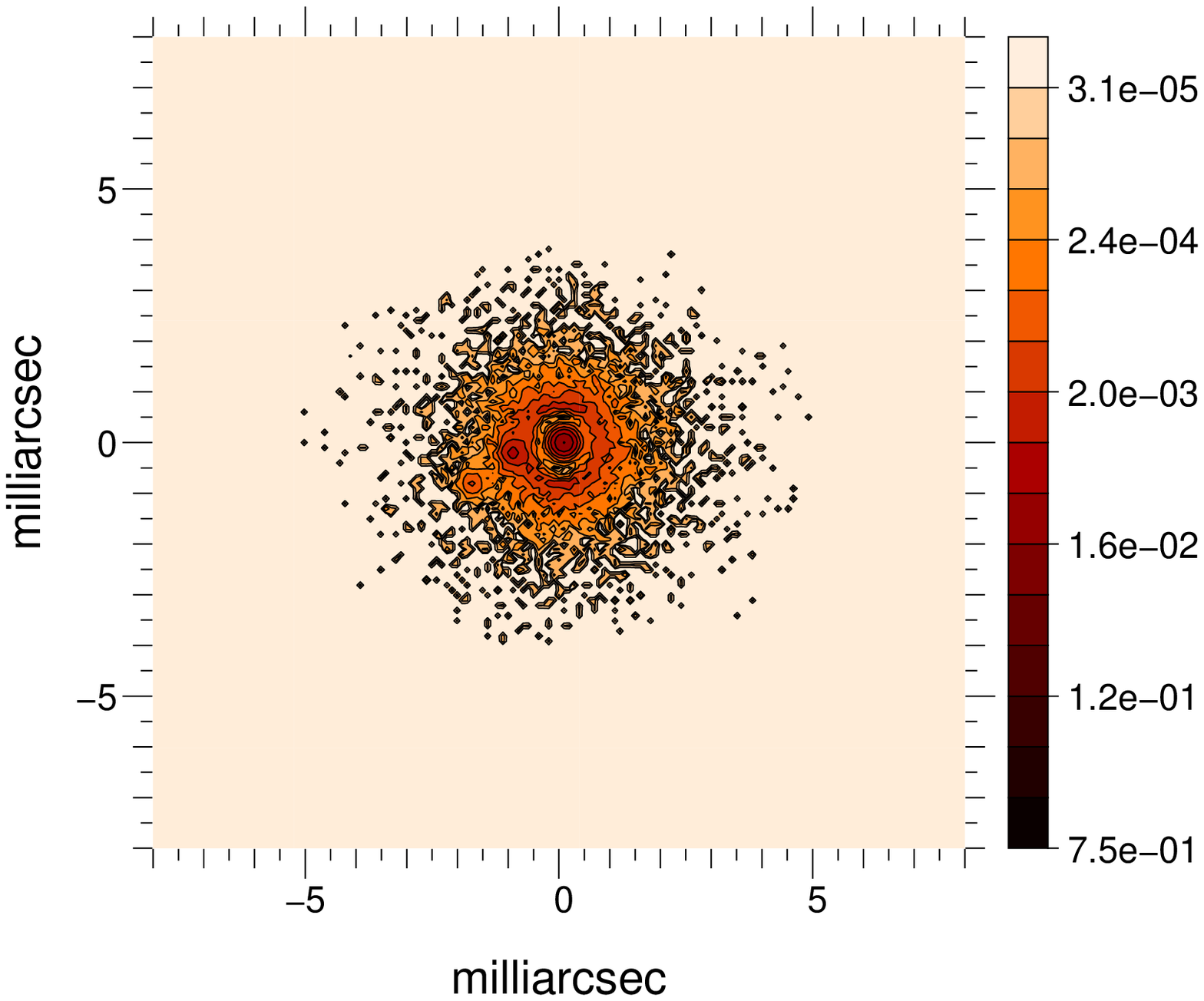}
  \includegraphics[height=4cm]{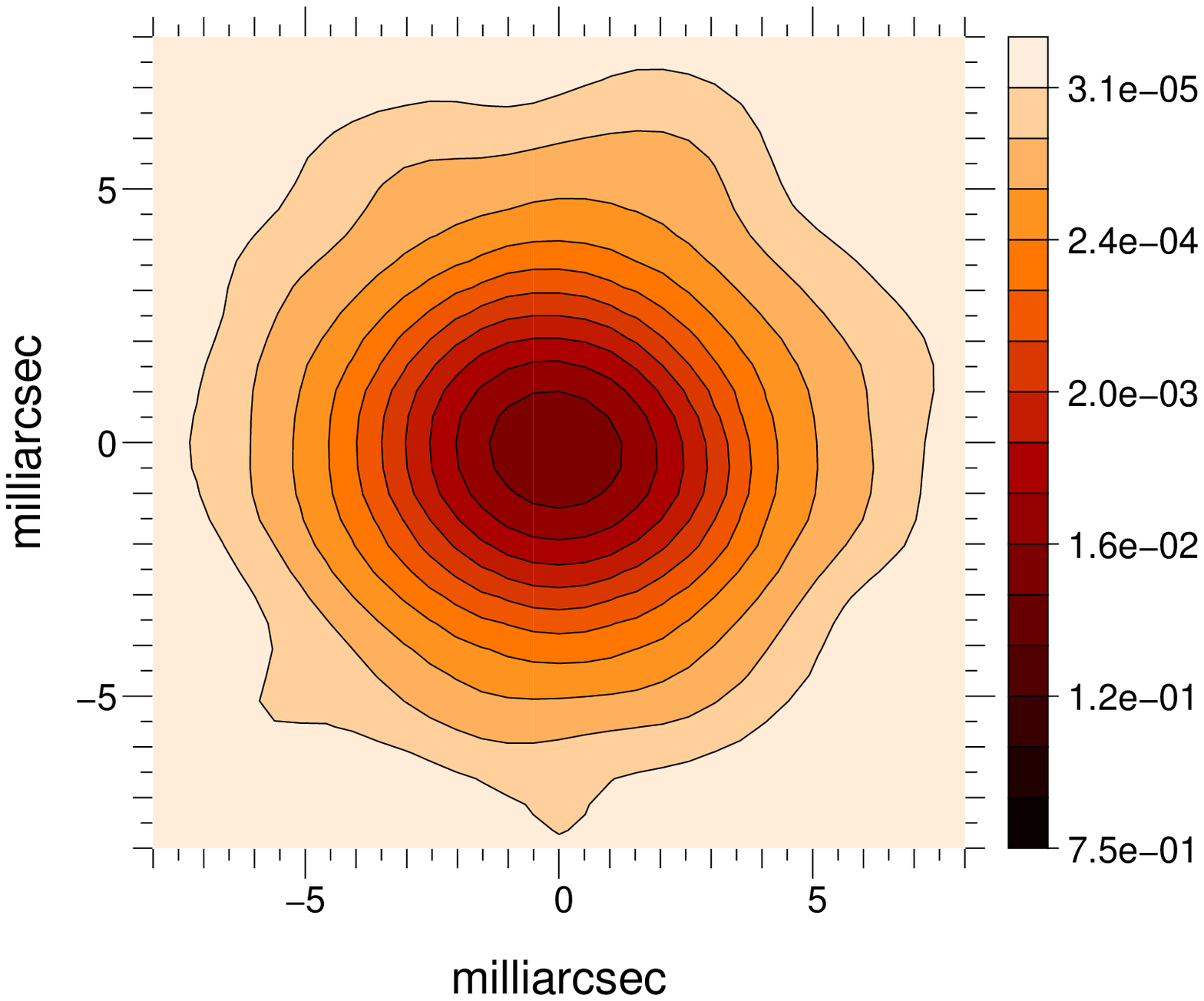}
\includegraphics[height=4cm]{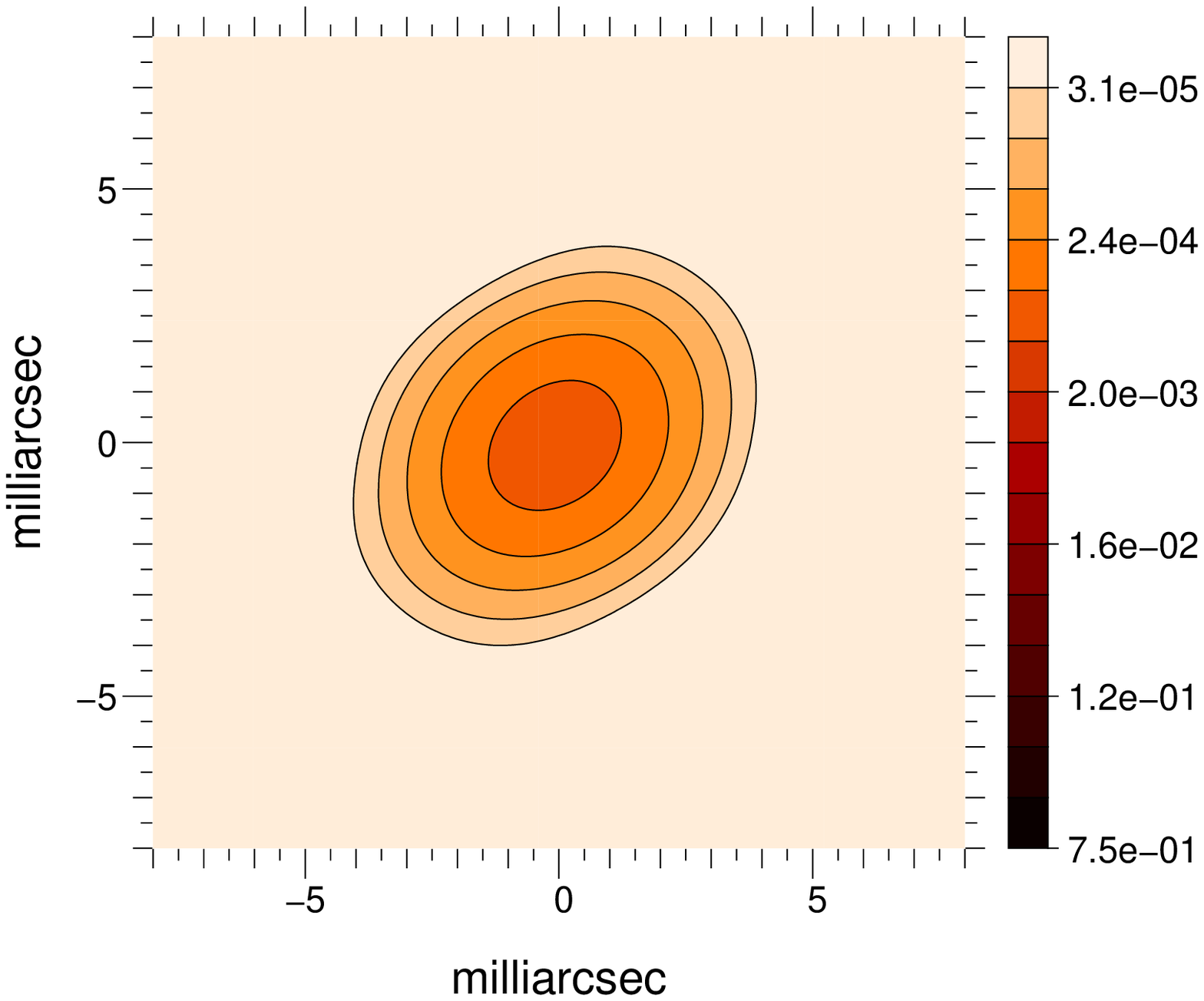}

   \end{tabular}
   \end{center}
   \caption[AGN]
   { \label{fig:example} A simulated AGN with jet component and torus, 0.1 mas/pixel sampling;  convolved image, 4 UT $\times$ 1 night configuration, 0.5 mas/pixel sampling, 4.45 mas FWHM resolution; {\sc aips} reconstruction 4 UT $\times$ 1 night configuration, 0.1 mas/pixel sampling.}
\end{figure}

\begin{table}[h!]
\begin{center}
\footnotesize
\caption {AGN. Diameter units are pixels.}
\begin{minipage}[c]{72mm}
\begin{tabular} {l | c c }

\hline

    & Image 4 UT  &         {\sc aips} 4 UT     \\
\hline

flux sublimation &  84.9\% & 97.0\% \\

flux torus & 15.1\% & 3.0\% \\

ratio & 5.6 & 32.3 \\

\hline

sublimation diameter & 50 & 45 \\

torus diameter & 260 &  - \\

\hline

SNR &  -  &  89 \\

\hline

\end{tabular}
\end{minipage}
\end{center}
\end{table}

















\begin{figure}[h!]
   \begin{center}
   \begin{tabular}{c}
\includegraphics[height=4cm]{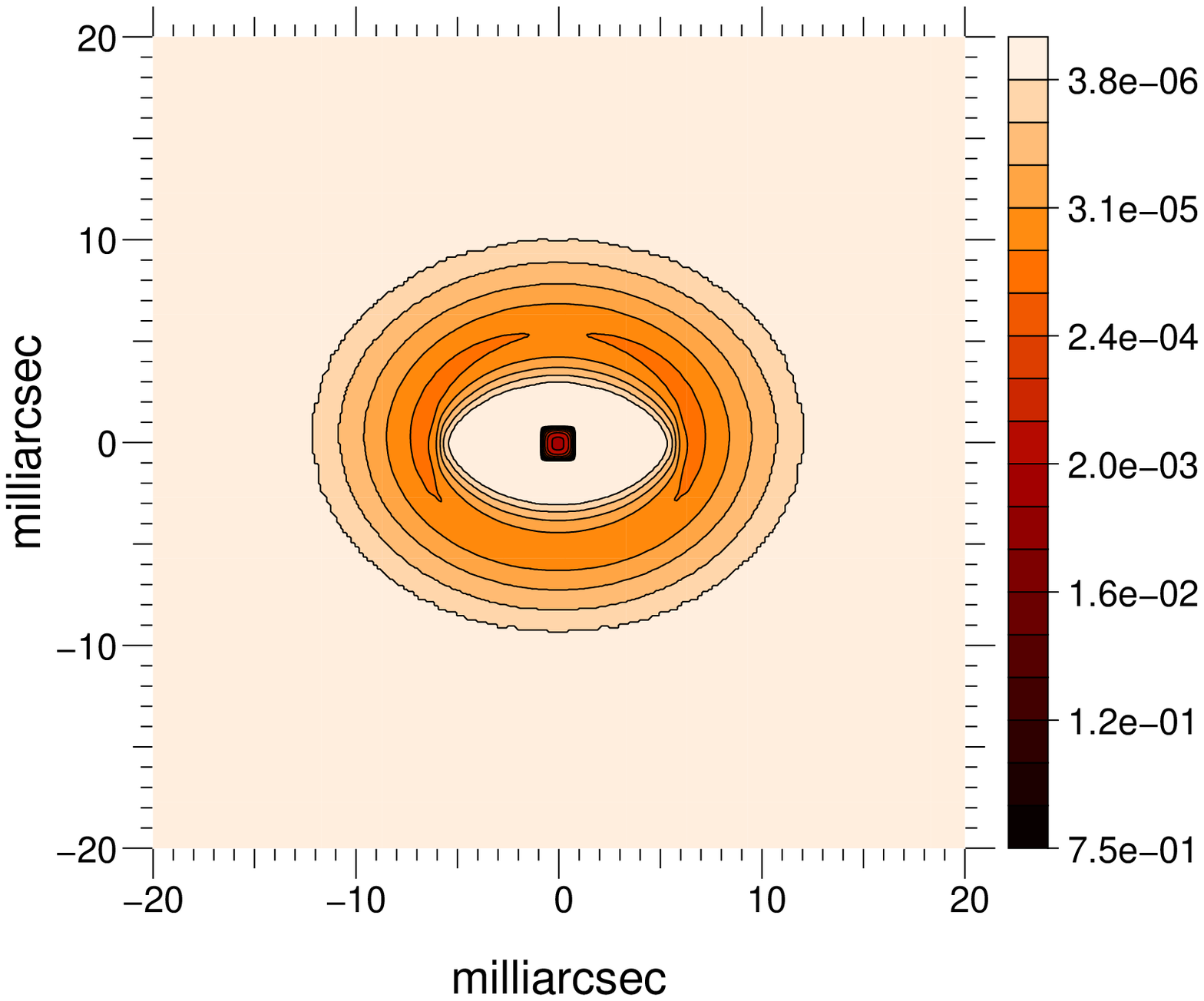}
\includegraphics[height=4cm]{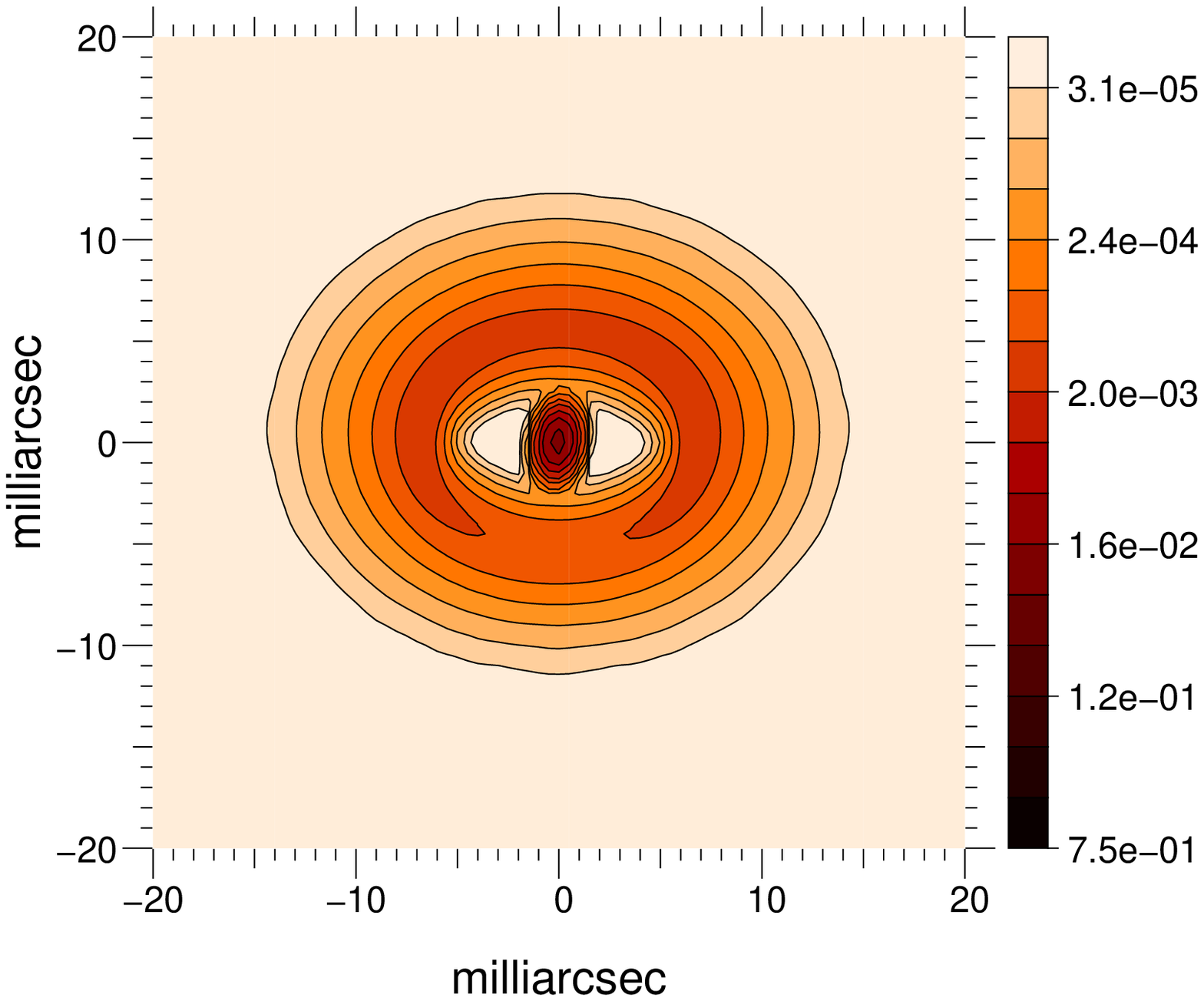}
\includegraphics[height=4cm]{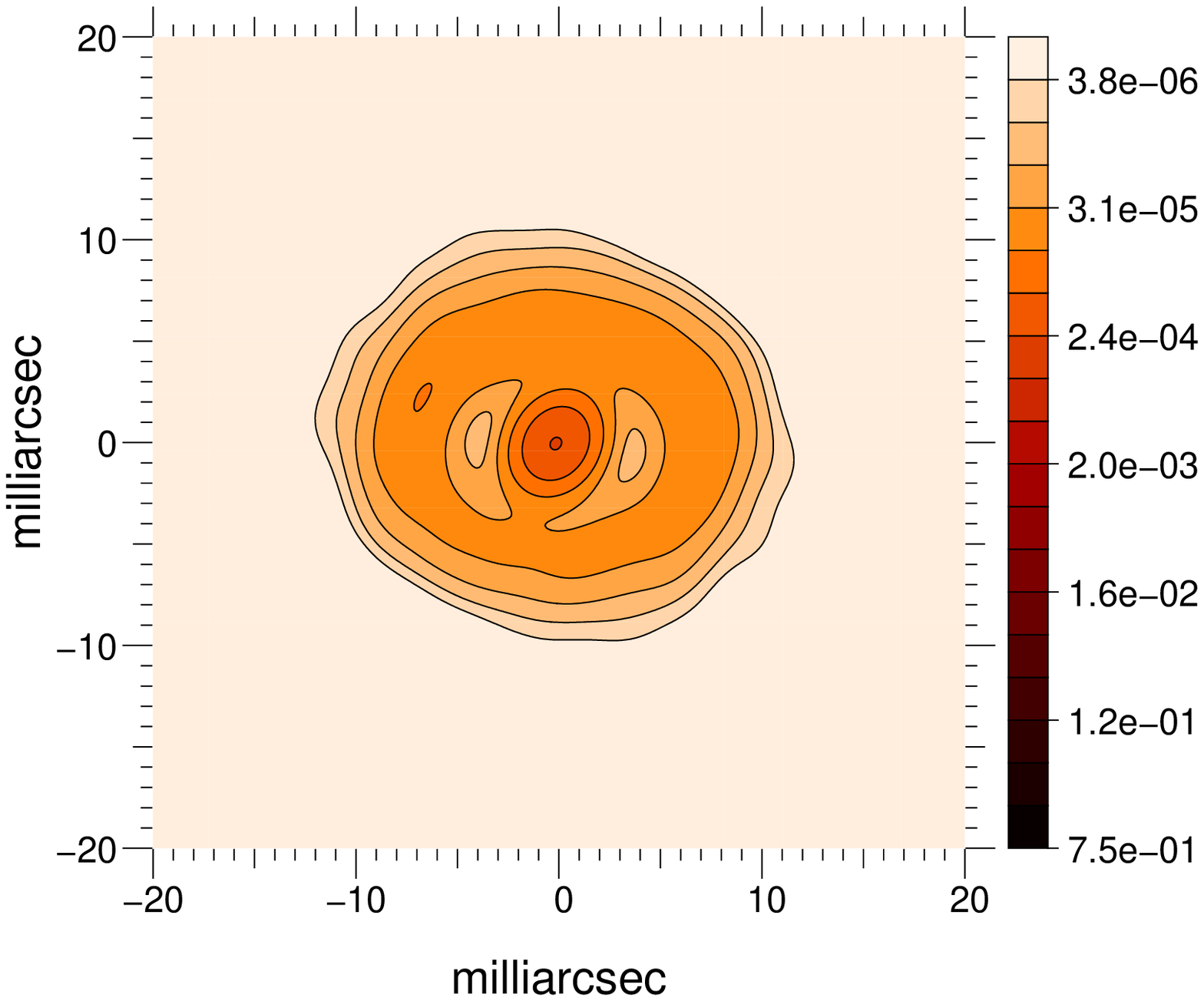}
   \end{tabular}
   \end{center}
\end{figure}

\begin{figure}[h!]
   \begin{center}
   \begin{tabular}{c}
\includegraphics[height=4cm]{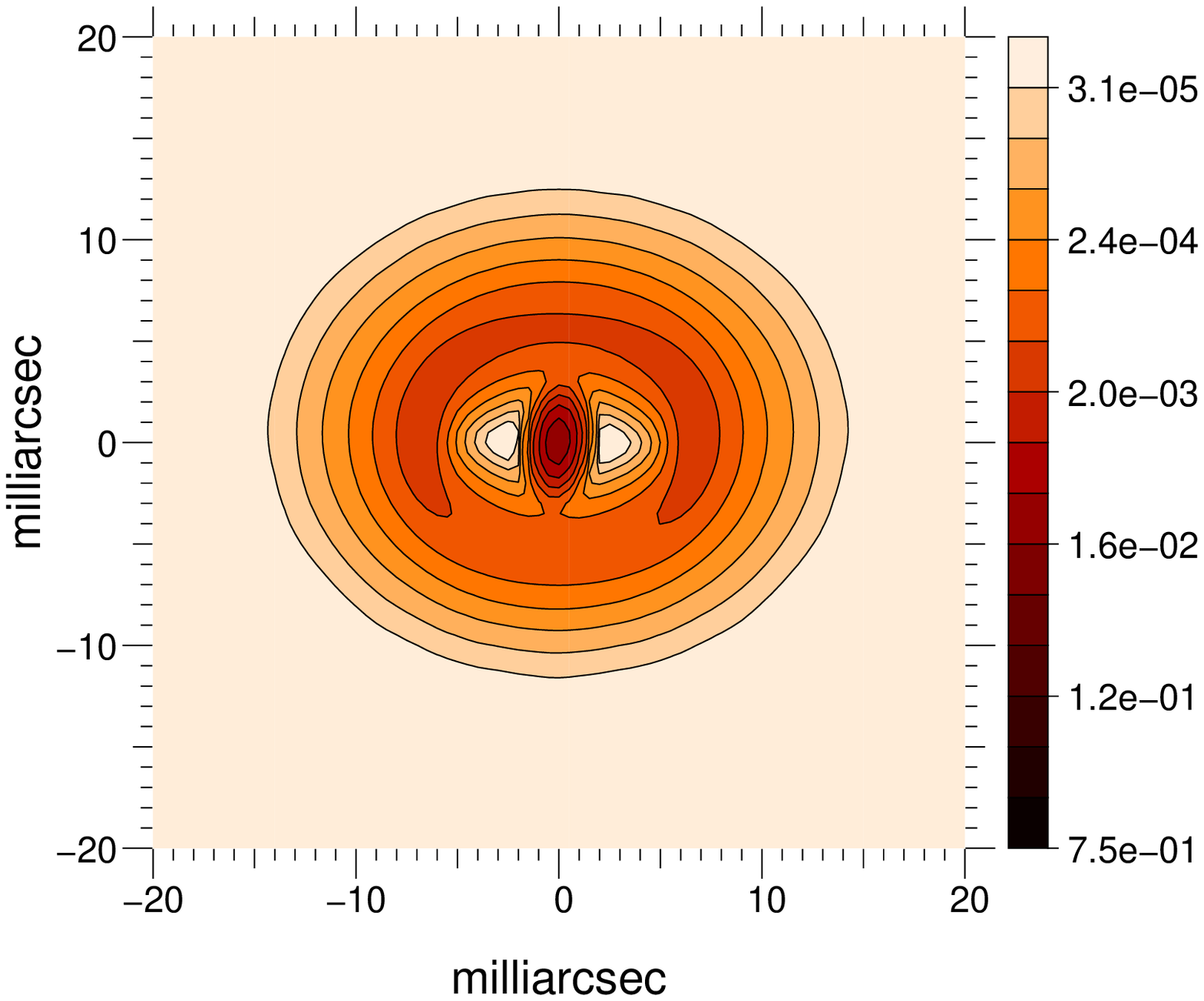}
\includegraphics[height=4cm]{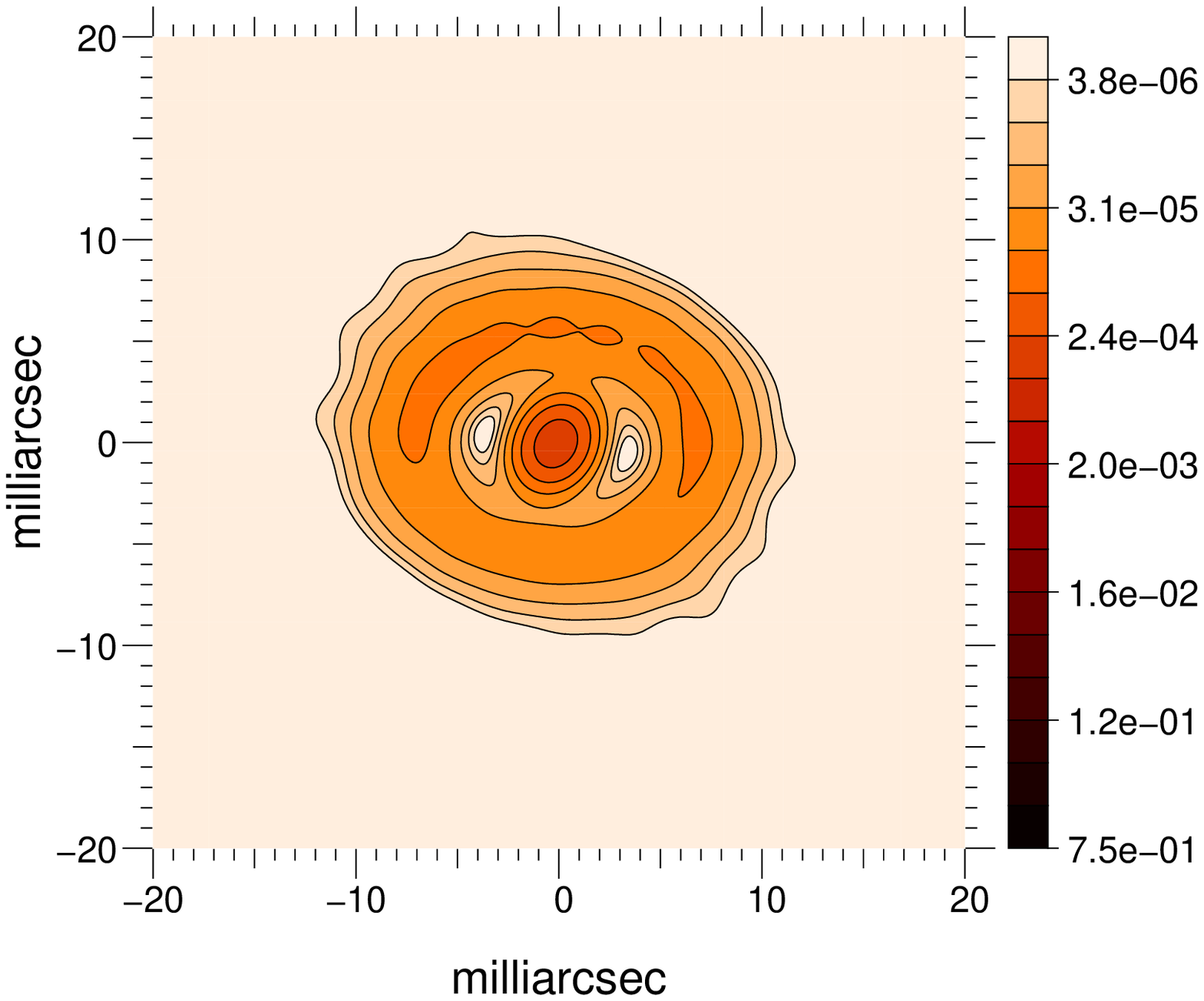}
   \end{tabular}
   \end{center}
   \caption[Evolved Star]
   { \label{fig:example} A low mass evolved star system with an outflow, 0.1 mas/pixel sampling; convolved image, 4 AT $\times$ 3 nights configuration, 0.5 mas/pixel sampling, 3.51 mas FWHM resolution; {\sc aips} reconstruction 4 AT $\times$ 3 nights configuration, 0.1 mas/pixel sampling; convolved image, 6 AT $\times$ 1 night configuration, 0.5 mas/pixel sampling, 5.17 mas FWHM resolution; {\sc aips} reconstruction 6 AT $\times$ 1 night configuration, 0.1 mas/pixel sampling.}
\end{figure}

\begin{table}[h!]
\begin{center}
\footnotesize
\caption{Evolved Star. Diameter units are pixels.}
\begin{minipage}[c]{125mm}
\begin{tabular} {l | c c c c }

\hline

      & Image     4 AT 3           &   {\sc aips} 4 AT 3  &    Image 6 AT & {\sc aips} 6 AT  \\

\hline

flux star   &    21.9\%         &   29.4\%   &   21.9\% & 28.7\%        \\

flux wind   &     78.1\%      &  70.6\%  &     78.1\%   & 71.3\%         \\

ratio star/wind &       0.3        &      0.4   &    0.3 & 0.4 \\

\hline

inner wind diameter &  50 $\times$ 35 &   50 $\times$ 35 &    50 $\times$ 35    &    50 $\times$ 35      \\

outer wind diameter &  100 $\times$ 85 & 100 $\times$ 85 & 100 $\times$ 85 & 100 $\times$ 85 \\

\hline

SNR                &       -           & 46              &       -          & 23 \\

\hline

\end{tabular}
\end{minipage}
\end{center}
\end{table}

\begin{figure}[h!]
   \begin{center}
   \begin{tabular}{c}
\includegraphics[height=4cm]{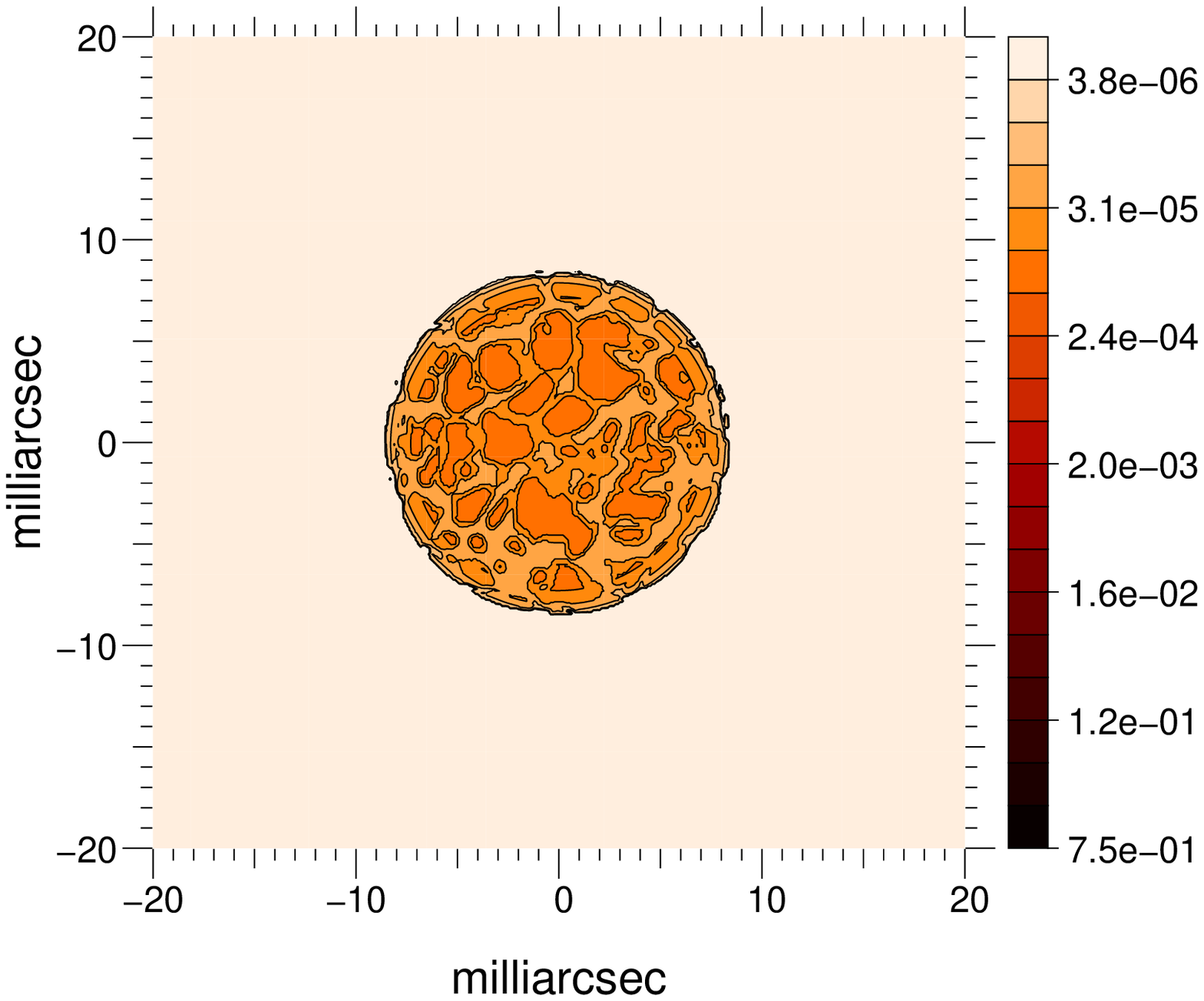}
\includegraphics[height=4cm]{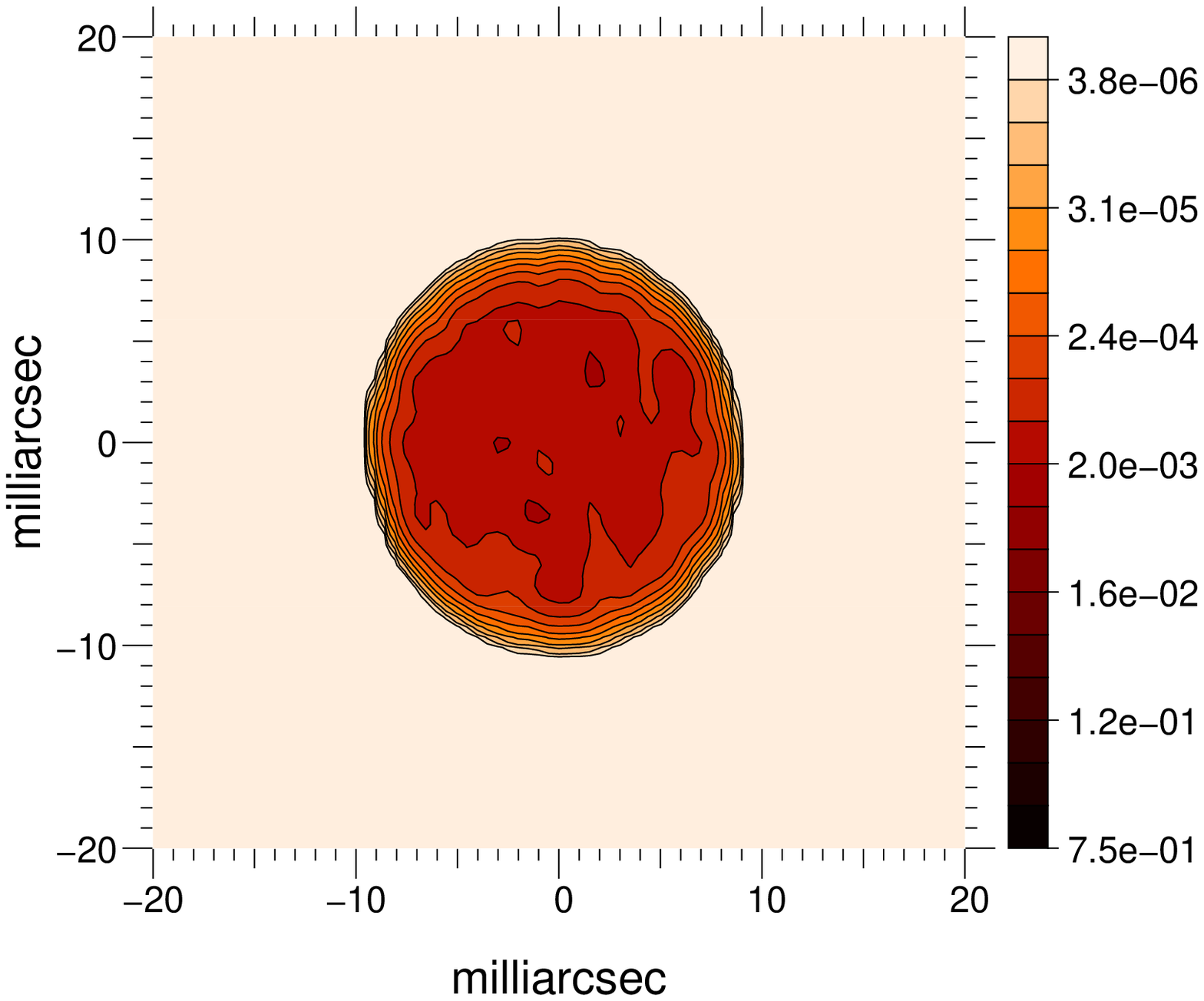}
\includegraphics[height=4cm]{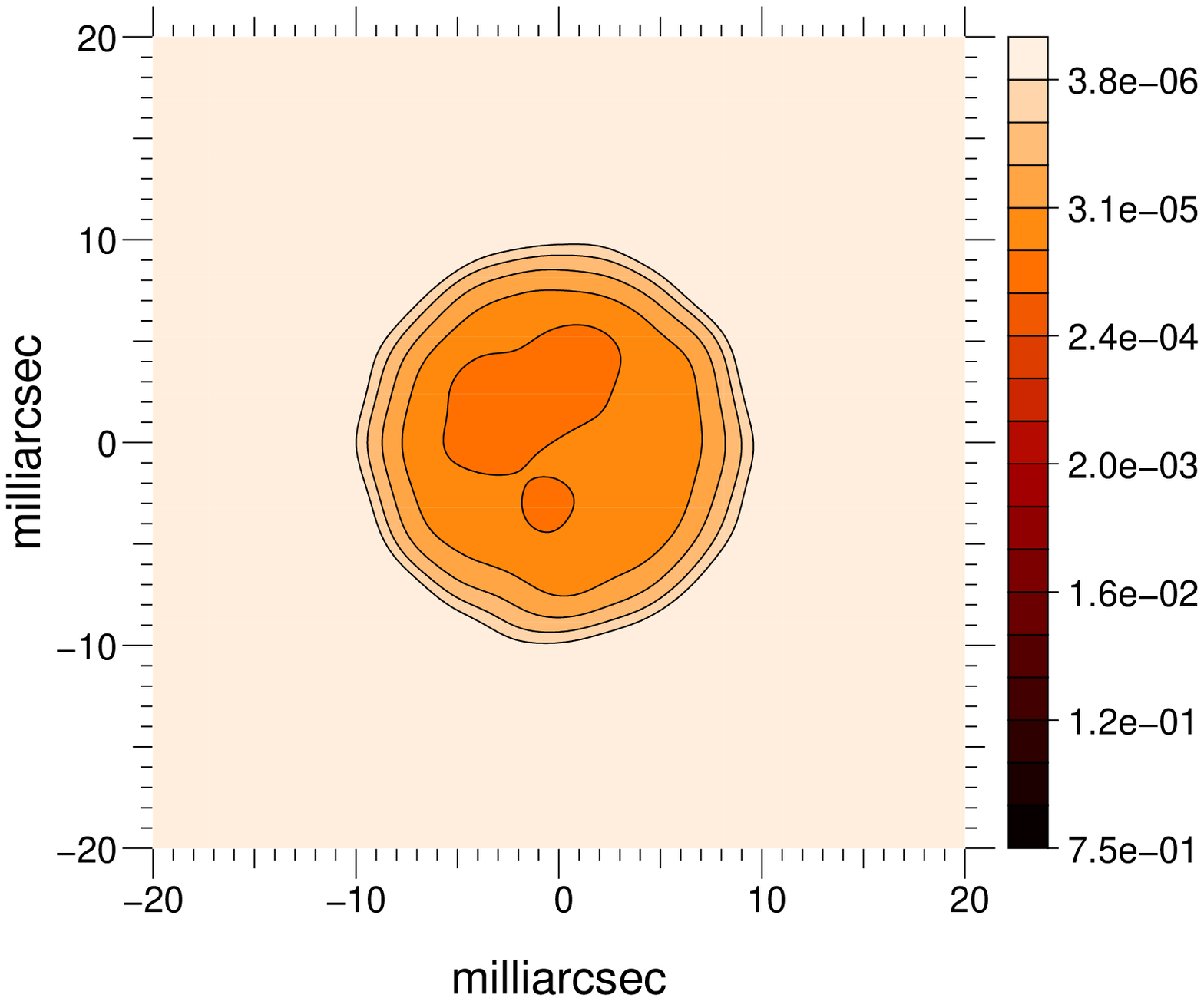}
   \end{tabular}
   \end{center}
\end{figure}

\begin{figure}[h!]
\begin{center}
   \begin{tabular}{c}
\includegraphics[height=4cm]{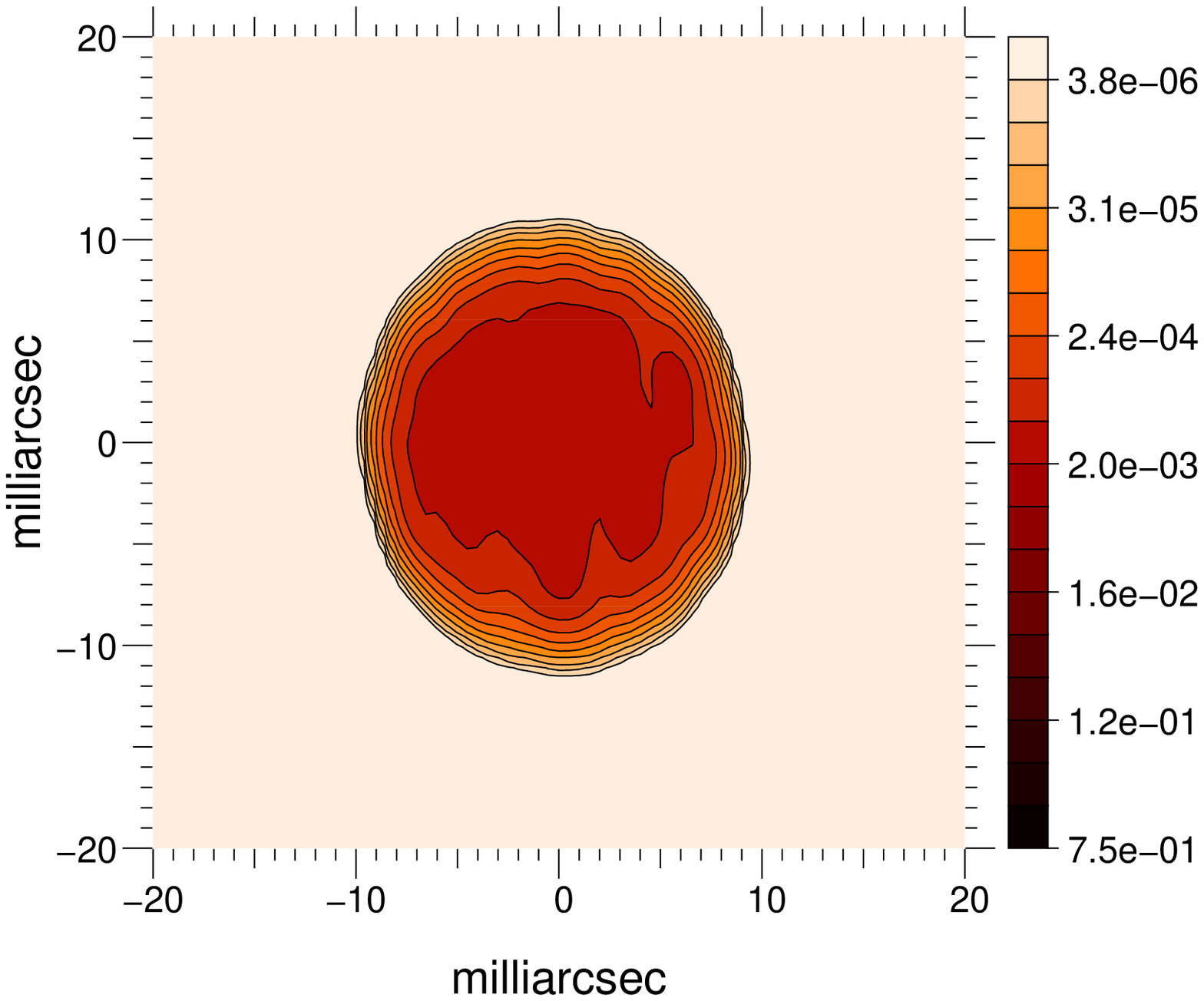}
\includegraphics[height=4cm]{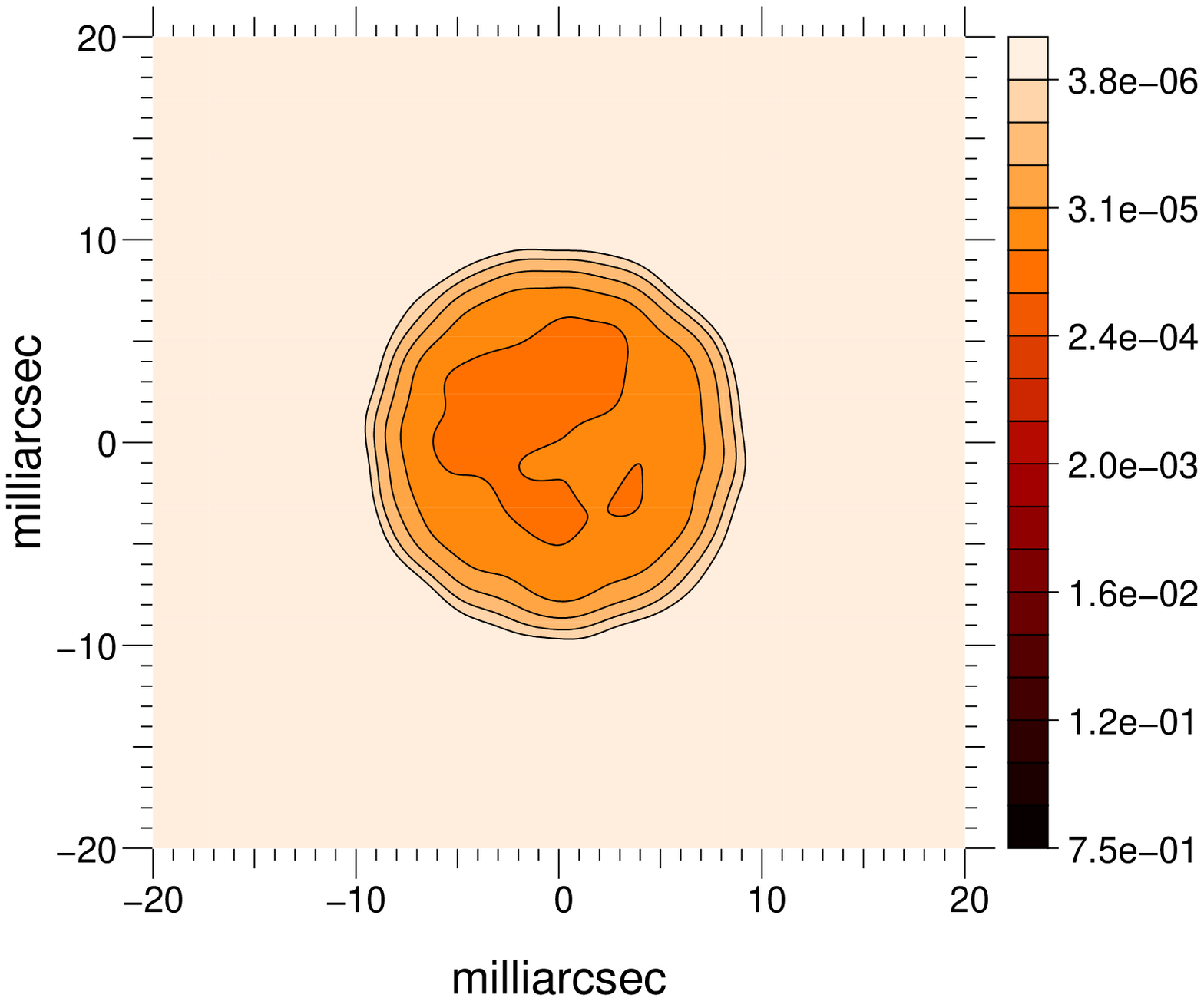}
\end{tabular}
\end{center}
   \caption[SG_surface]
 { \label{fig:example} A simulated stellar surface of a M giant, 0.1 mas/pixel sampling; convolved image, 4 AT $\times$ 3 nights configuration, 0.5 mas/pixel sampling, 3.51 mas FWHM resolution; {\sc aips} reconstruction 4 AT $\times$ 3 nights configuration, 0.1 mas/pixel sampling, $SNR$ = 34; convolved image, 6 AT $\times$ 1 night configuration, 0.5 mas/pixel sampling, 5.17 mas FWHM resolution; {\sc aips} reconstruction 6 AT $\times$ 1 night configuration, 0.1 mas/pixel sampling, $SNR$ = 24.}
\end{figure}










\begin{figure}[h!]
   \begin{center}
   \begin{tabular}{c}
\includegraphics[height=4cm]{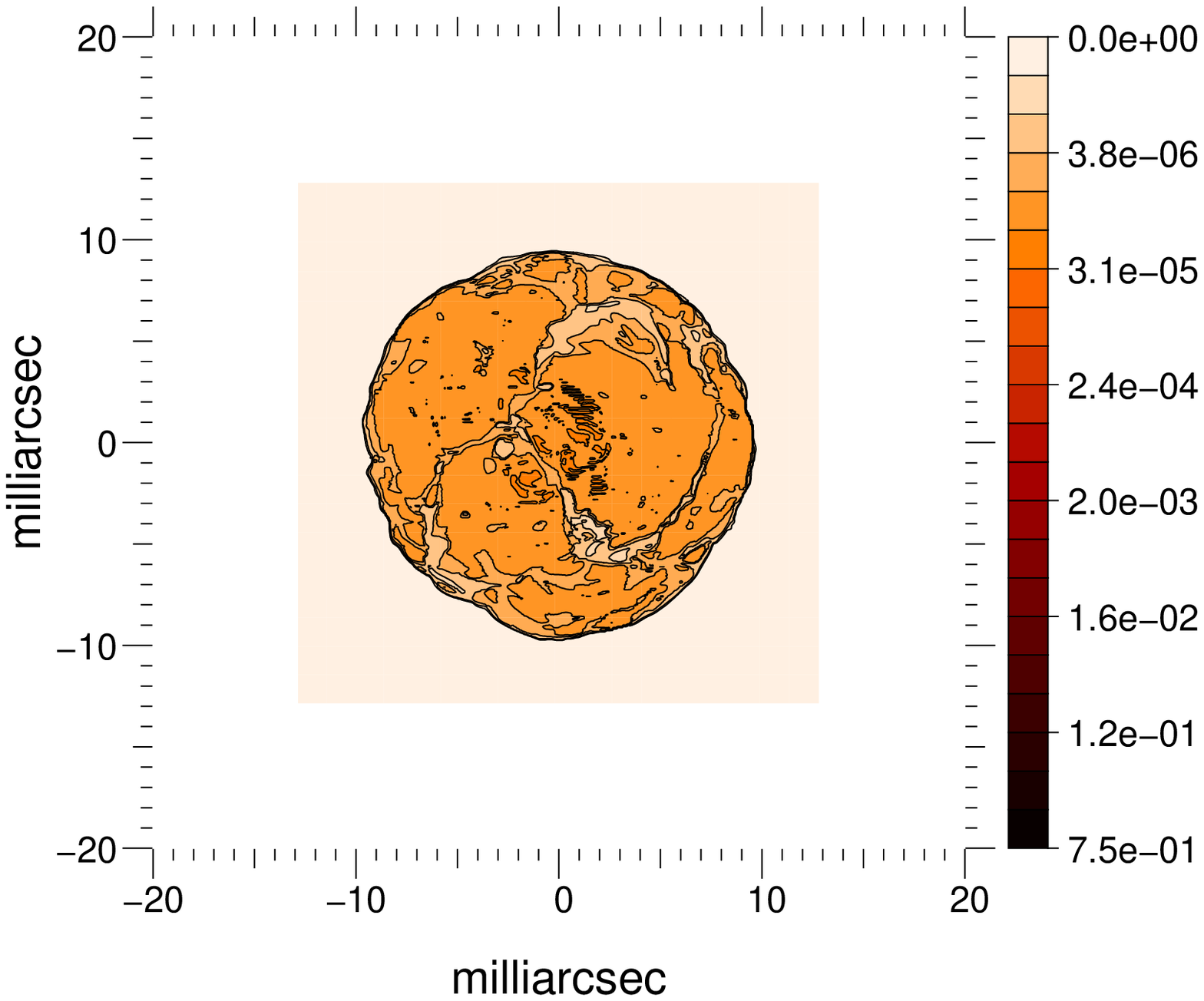}
\includegraphics[height=4cm]{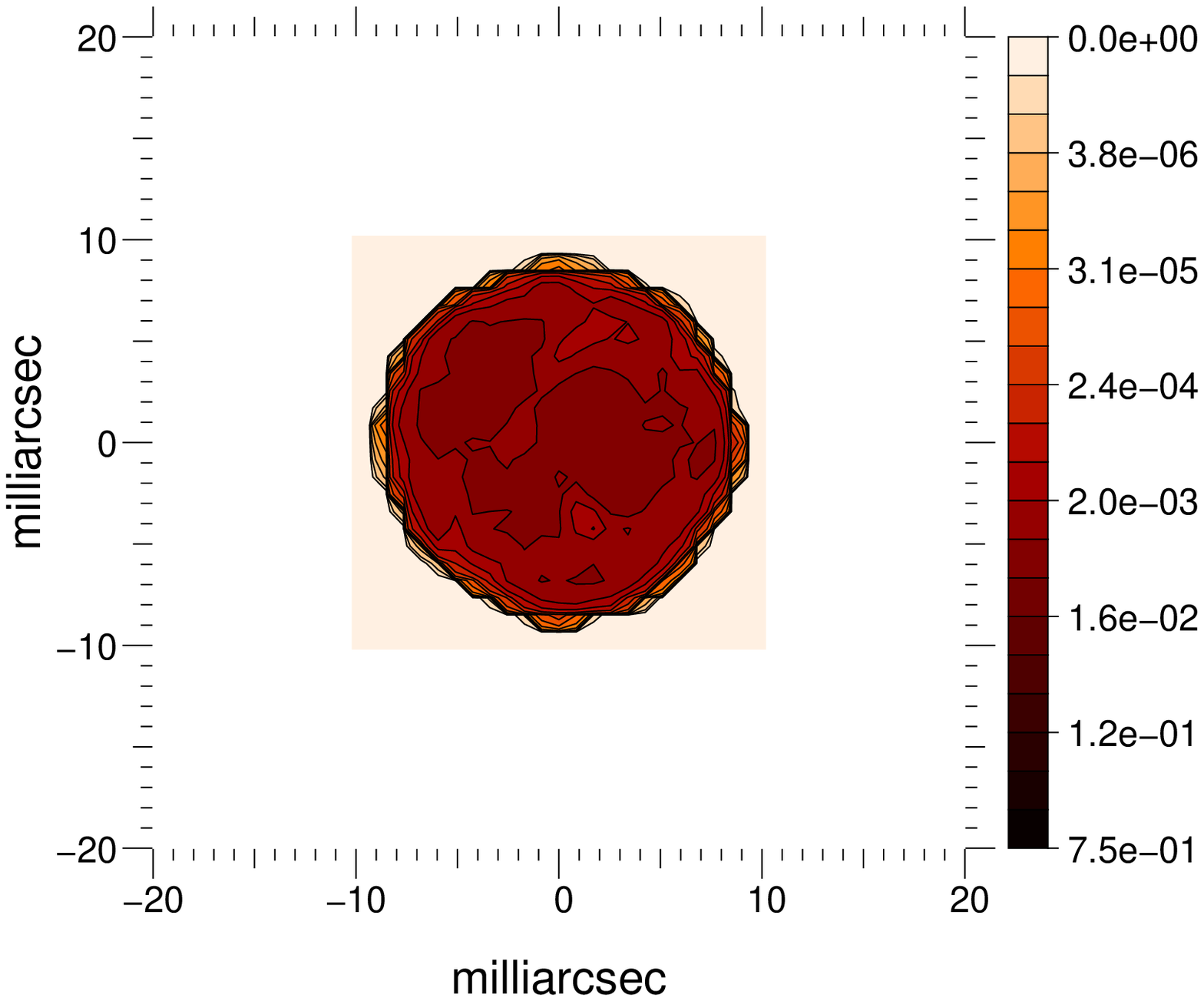}
\includegraphics[height=4cm]{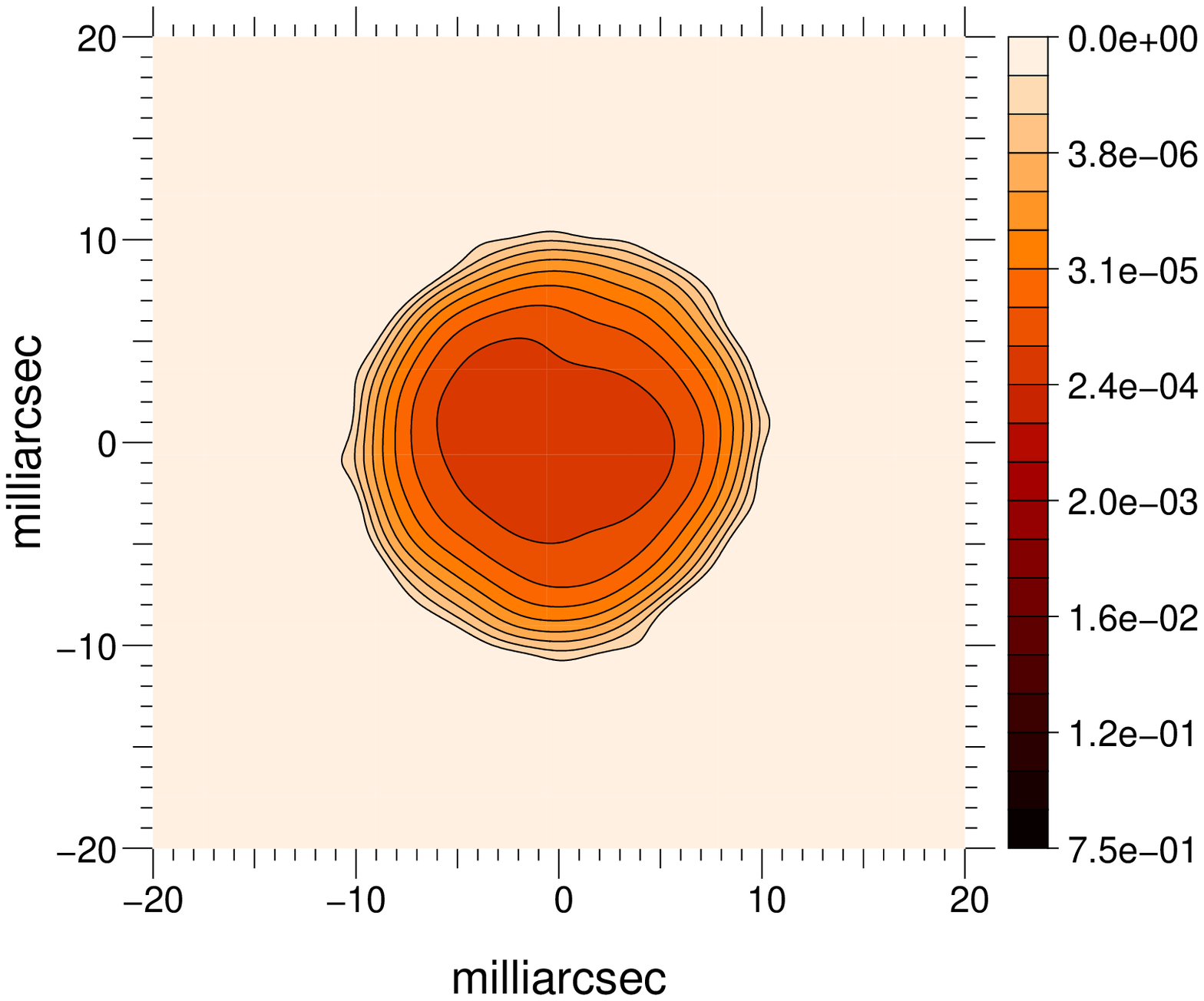}
\end{tabular}
\end{center}
   \caption[SG_surface2]
   { \label{fig:example} A simulated stellar surface of a supergiant, 0.1 mas/pixel sampling; convolved image, 6 AT $\times$ 1 night configuration, 0.5 mas/pixel sampling, 5.17 mas FWHM resolution; {\sc aips} reconstruction 6 AT $\times$ 1 night configuration, 0.1 mas/pixel sampling, $SNR$ = 216.}
\end{figure}








\begin{figure}[h!]
   \begin{center}
   \begin{tabular}{c}
\includegraphics[height=4cm]{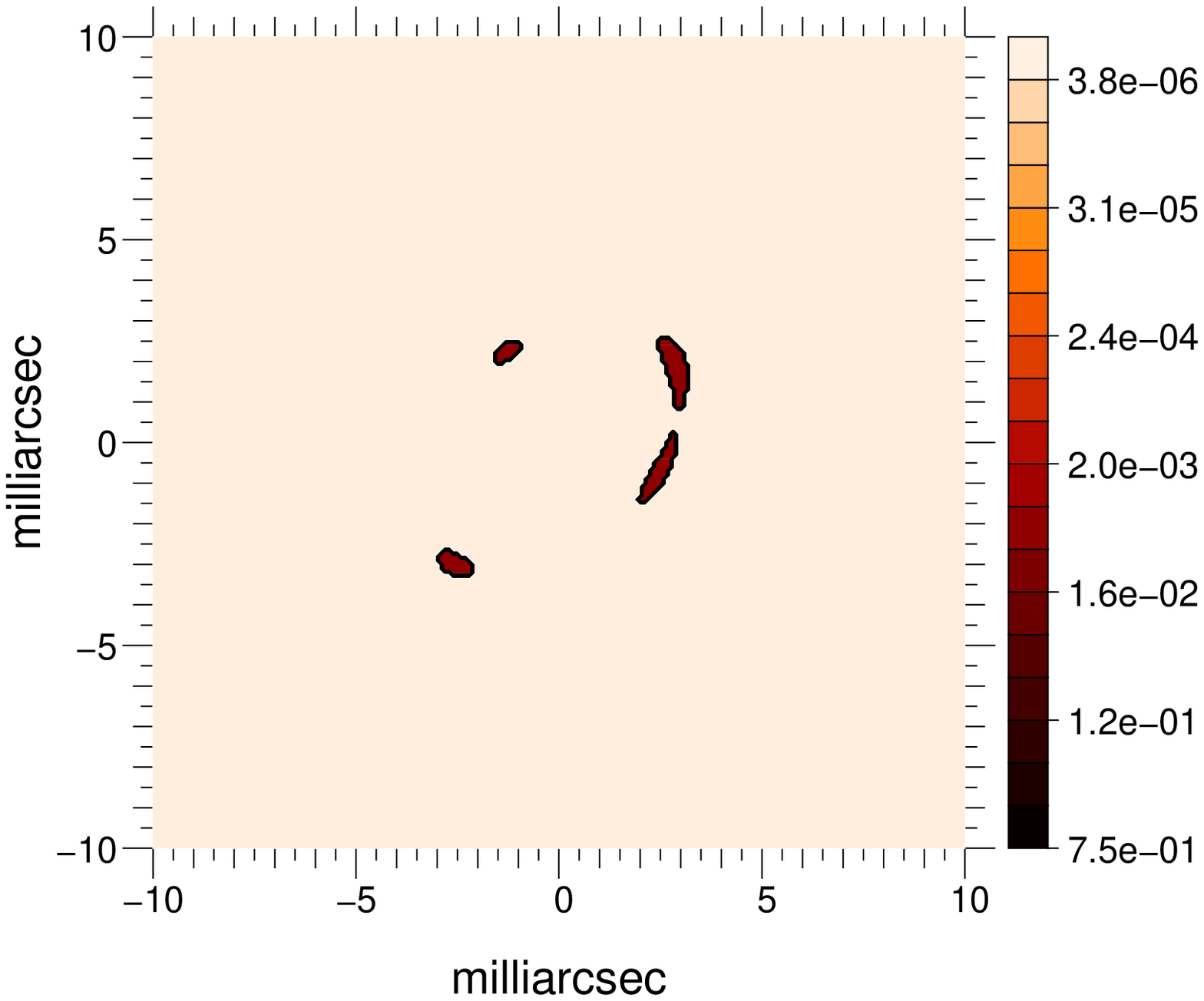}
\includegraphics[height=4cm]{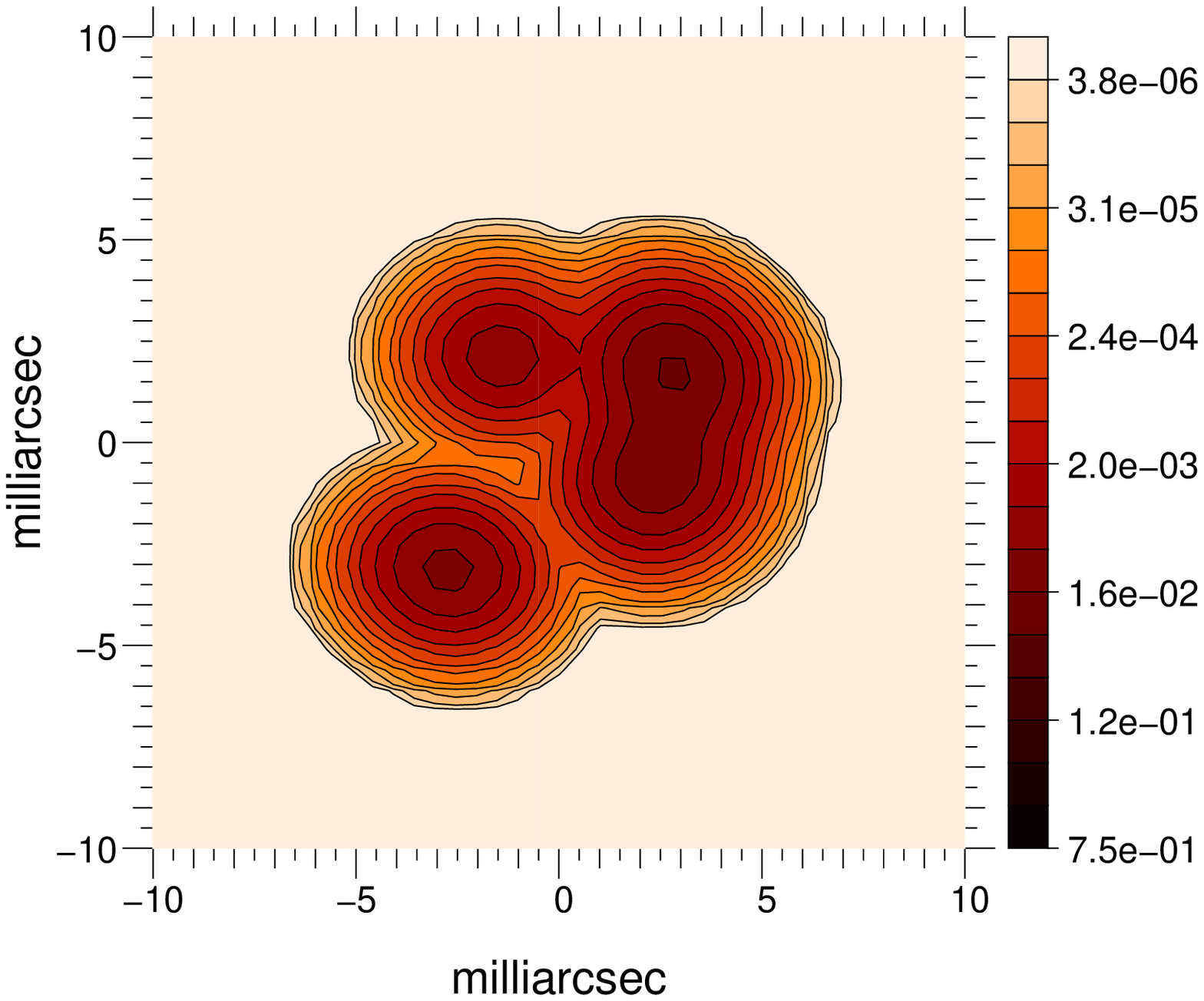}
\includegraphics[height=4cm]{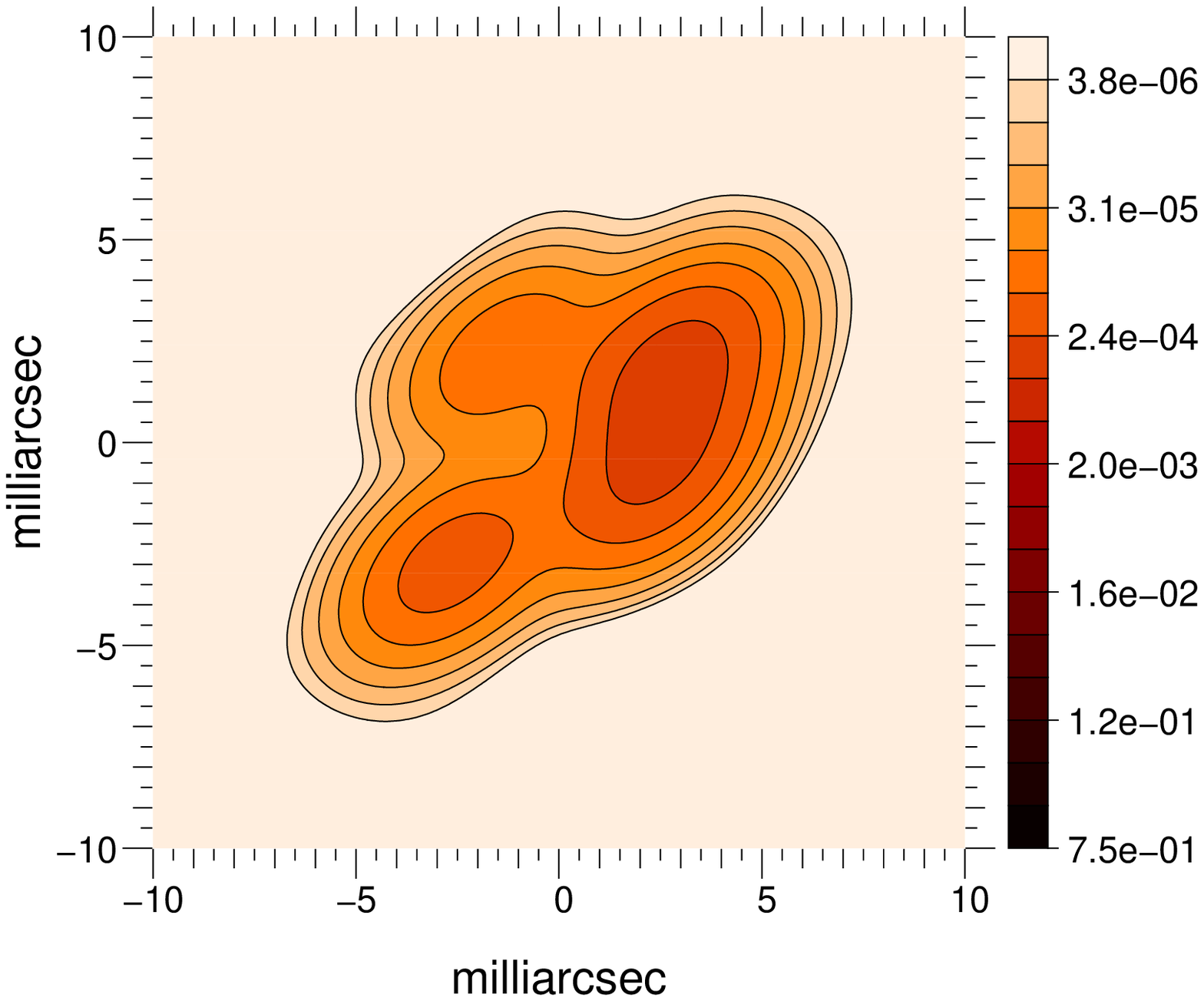}
 \end{tabular}
   \end{center}
   \caption[microlensing]
   { \label{fig:example} A simulated microlensing event, 0.1 mas/pixel sampling; convolved image, 4 UT $\times$ 1 night configuration, 0.5/pixel sampling, 4.45 mas FWHM resolution; {\sc aips} reconstruction 4 UT $\times$ 1 nights configuration, 0.1 mas/pixel sampling.}
\end{figure}

\begin{table}[h!]
\begin{center}
\footnotesize
\caption{Microlensing. Distance units are pixels.}
\begin{minipage}[c]{60mm}
\begin{tabular} {l | c c }

\hline
    &   Image   4 UT    &   {\sc aips} 4 UT    \\

\hline

A flux  &       12.8\% & 12.9\%  \\


B flux  &       19.2\% & 19.6\% \\


C flux & 68.0\%  & 67.5\% \\

ratio C/A &     5.3 &   5.2 \\

ratio C/B &     3.5 &    3.4   \\

\hline

distance AC & 45 & 45 \\

distance BC & 70 & 70 \\

distance AB & 50 & 50 \\

\hline 

SNR         & -  & 99 \\

\hline




\hline

\end{tabular}
\end{minipage}
\end{center}
\end{table}





\begin{figure}[h!]
   \begin{center}
   \begin{tabular}{c}
 \includegraphics[height=4cm]{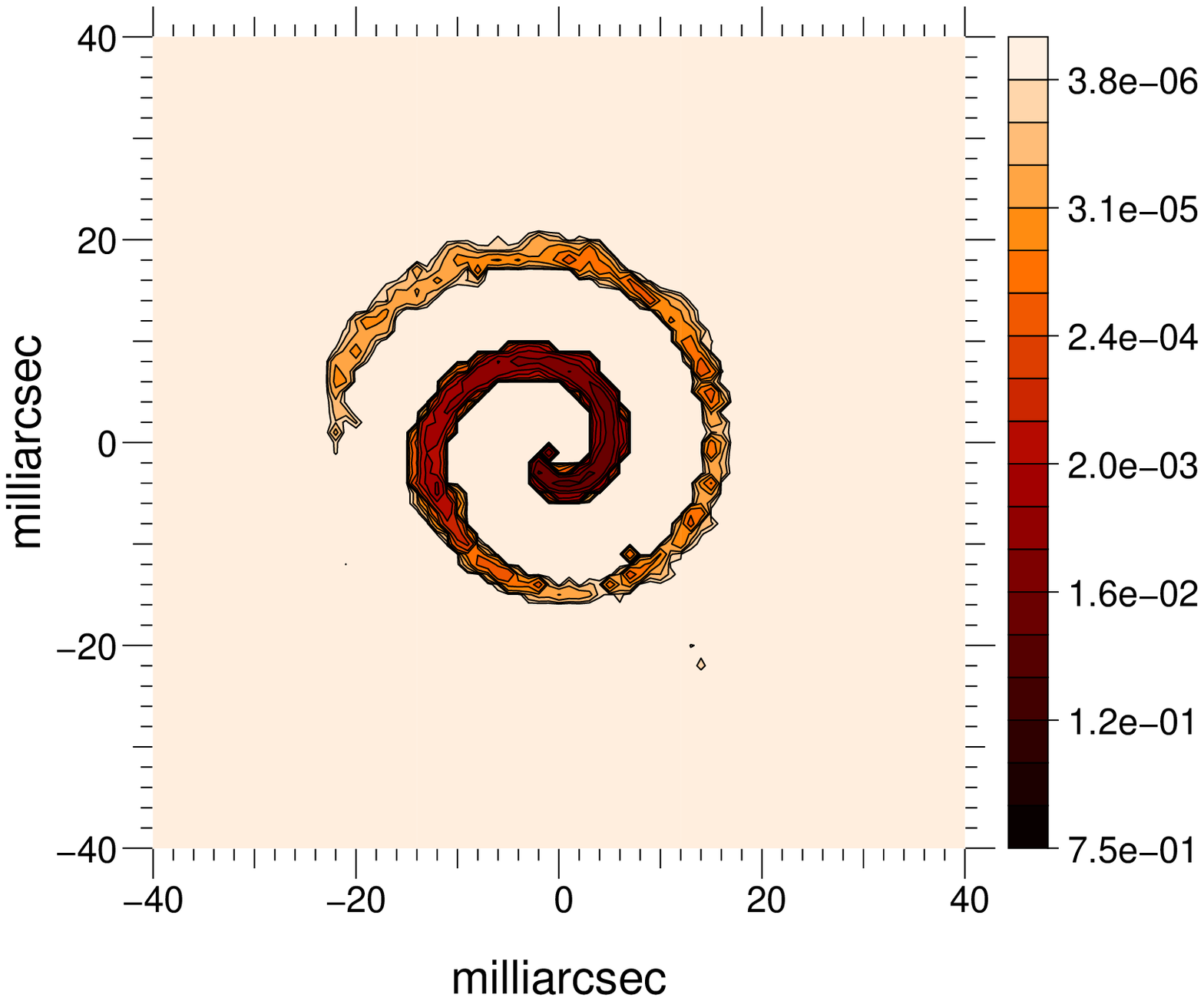}
 \includegraphics[height=4cm]{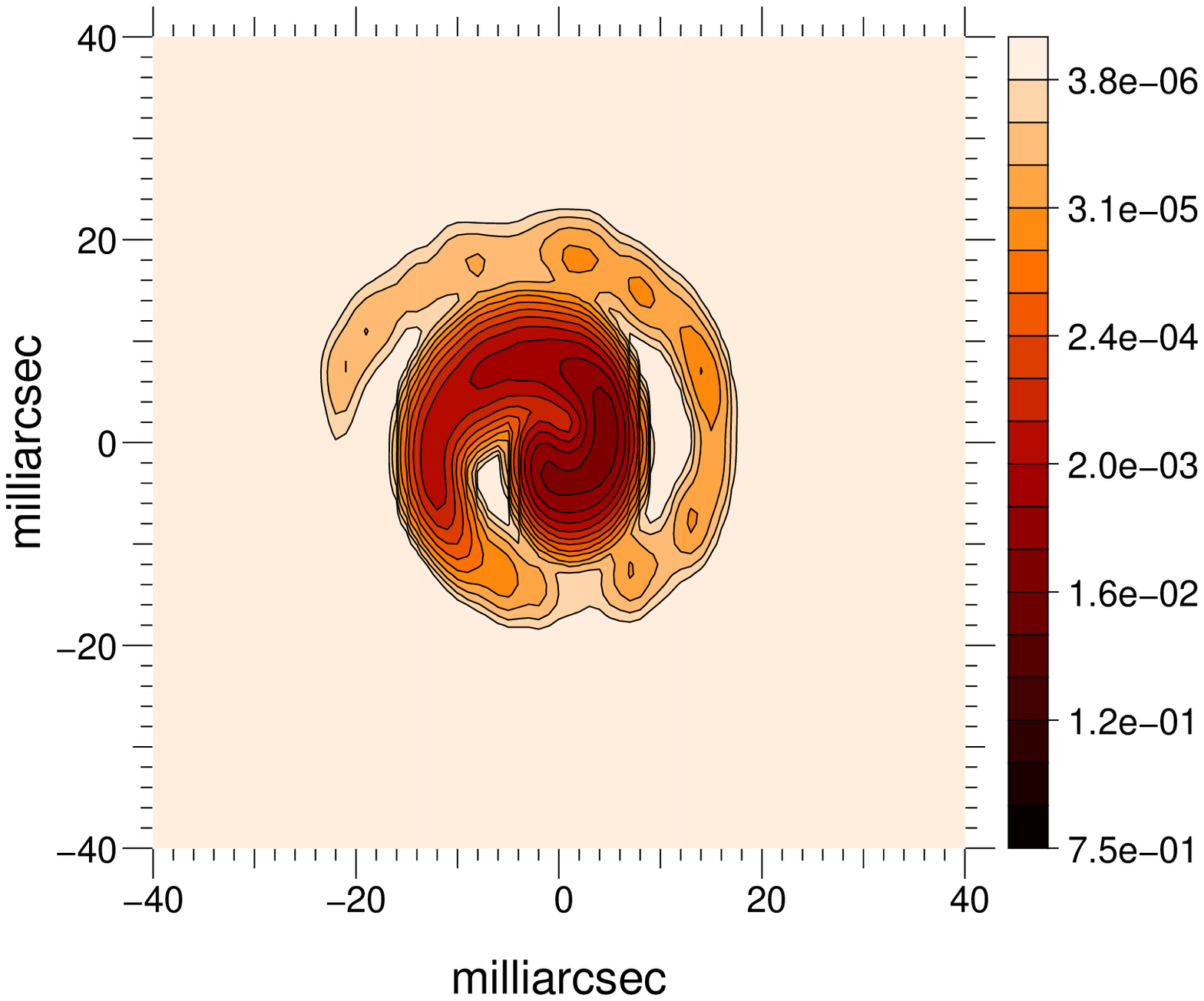}
 \includegraphics[height=4cm]{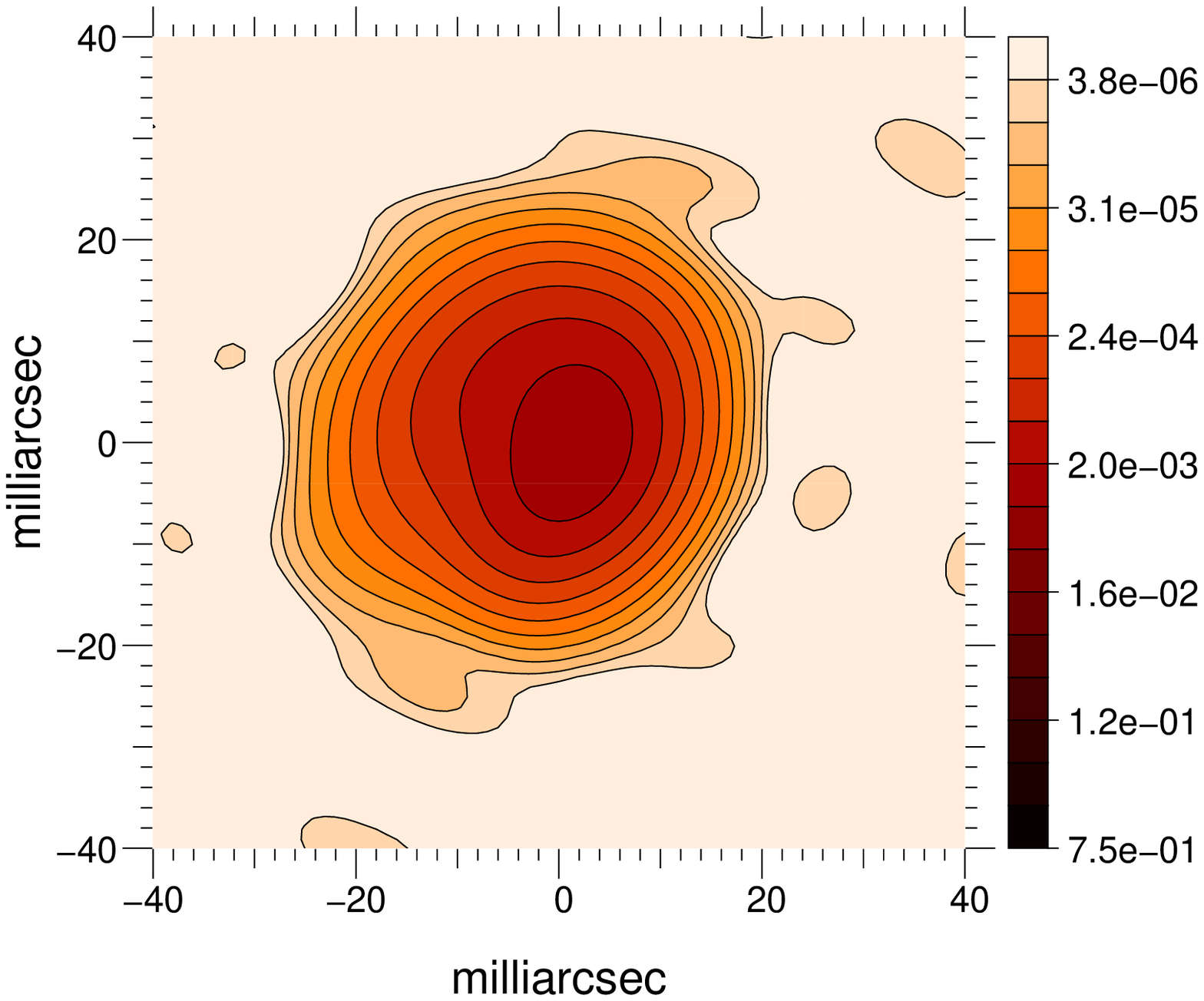}
 \end{tabular}
   \end{center}
\end{figure}

\begin{figure}[h!]
   \begin{center}
   \begin{tabular}{c}
\includegraphics[height=4cm]{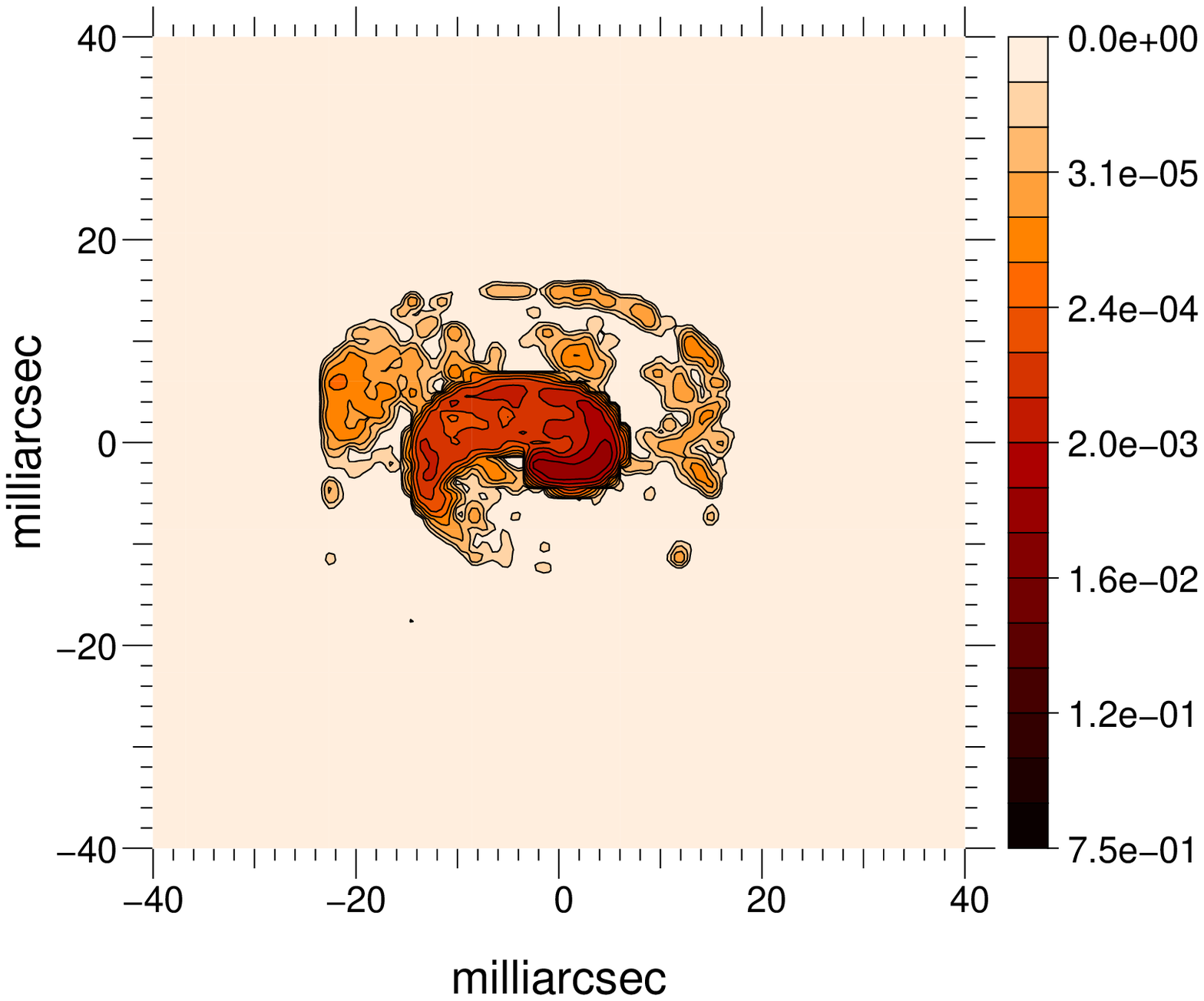}
\includegraphics[height=4cm]{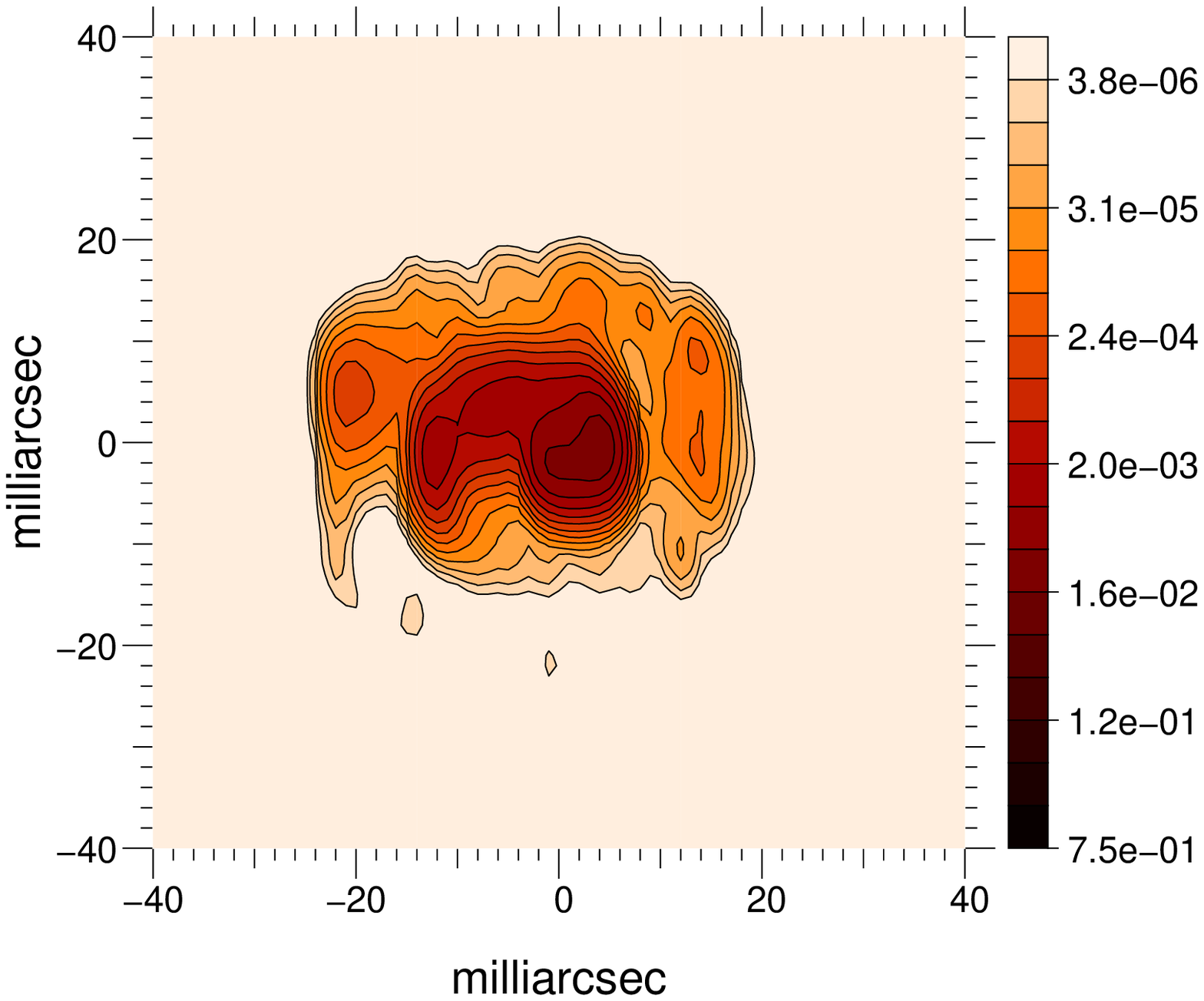}
\includegraphics[height=4cm]{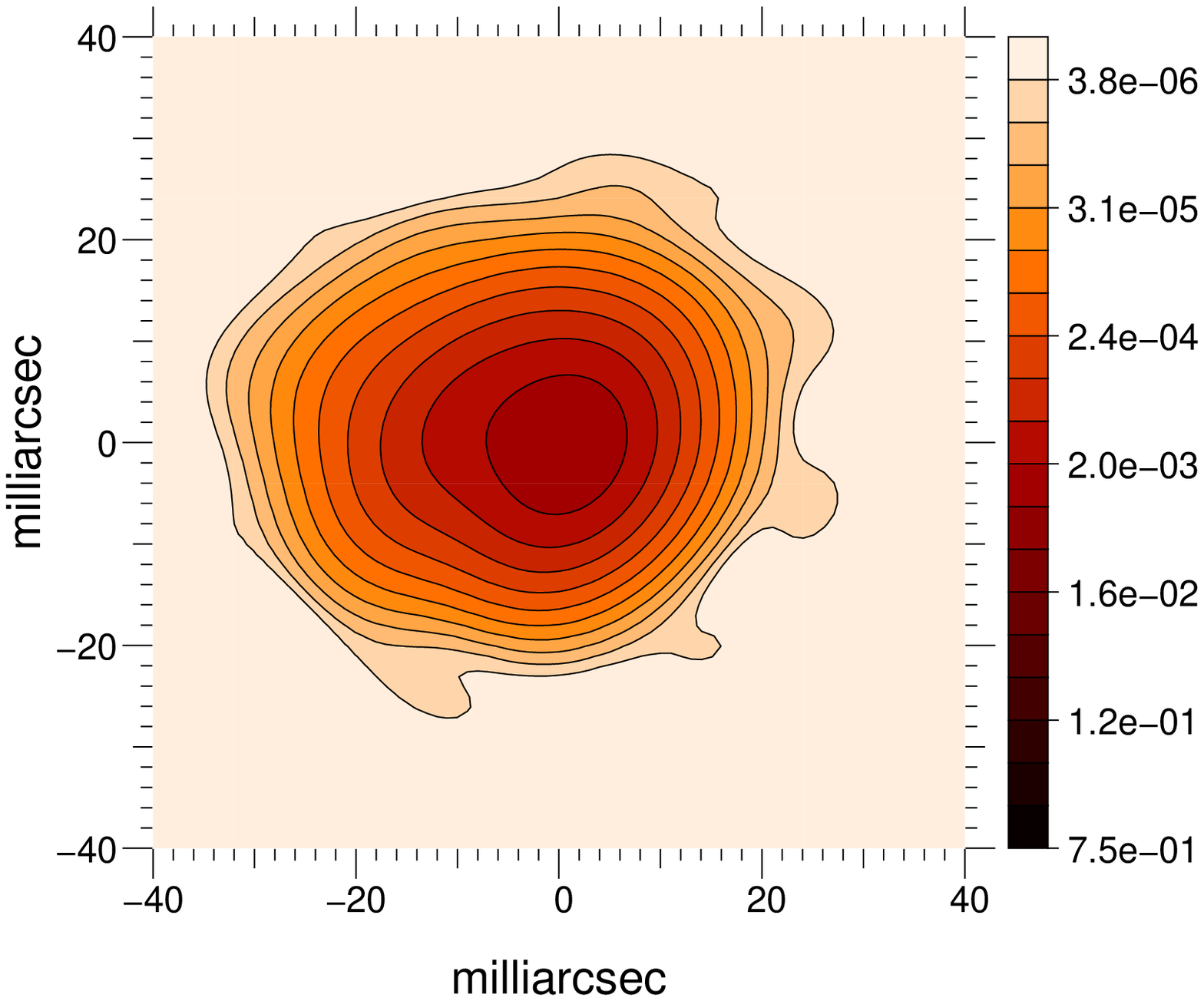}
   \end{tabular}
   \end{center}
   \caption[microlensing]
   { \label{fig:example} A simulated image of the pinwheel nebula at 0 degree inclination, 1.0 mas/pixel sampling; convolved image, 1.0 mas/pixel sampling, 5.17 FWHM, 6 AT $\times$ 1 night configuration; {\sc aips} reconstruction 6 AT $\times$ 1 night configuration, 0.1 mas/pixel sampling. A simulated image of the pinwheel nebula at 60 degree inclination, 1.0 mas/pixel sampling; convolved image, 1.0 mas/pixel sampling, 5.17 FWHM, 6 AT $\times$ 1 night configuration;  {\sc aips} reconstruction 6 AT $\times$ 1 night configuration, smoothness regularization, 0.1 mas/pixel sampling.}
\end{figure}

\begin{table}[h!]
\begin{center}
\footnotesize
\caption{Pinwheel}
\begin{minipage}[c]{85mm}
\begin{tabular} {l | c c c c c c c }

\hline

    & Angle &   Image  6 AT     & {\sc aips} 6 AT & SNR \\

\hline

inner spiral  & 0 deg &   24 $\times$ 30     &   24 $\times$ 24  & 281 \\

inner spiral & 60 deg &   24 $\times$ 24     &   24 $\times$ 22  & 404 \\

\hline

\end{tabular}
\end{minipage}
\end{center}
\end{table}

\begin{figure}[h!]
   \begin{center}
   \begin{tabular}{c}
\includegraphics[height=4cm]{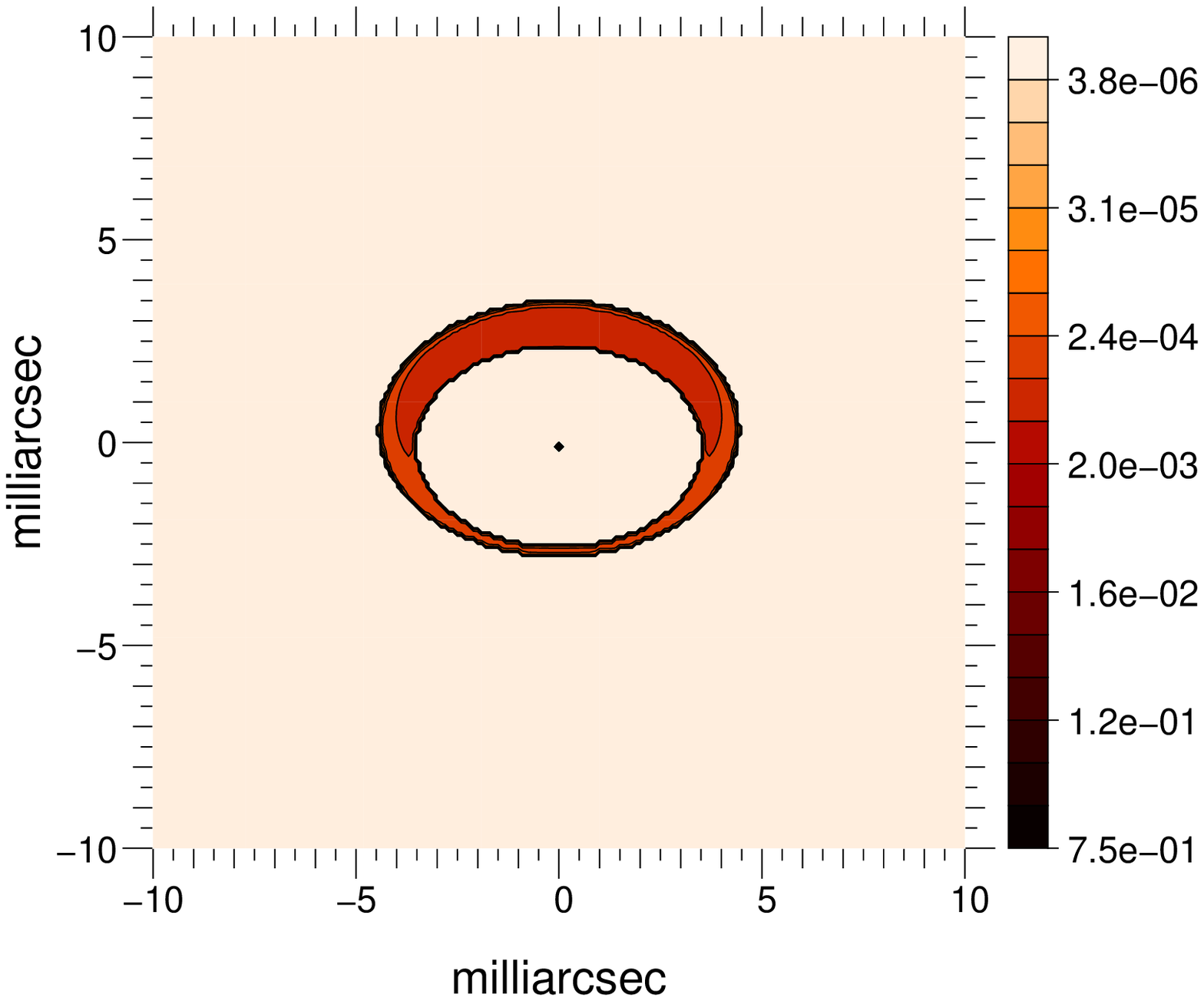}
\includegraphics[height=4cm]{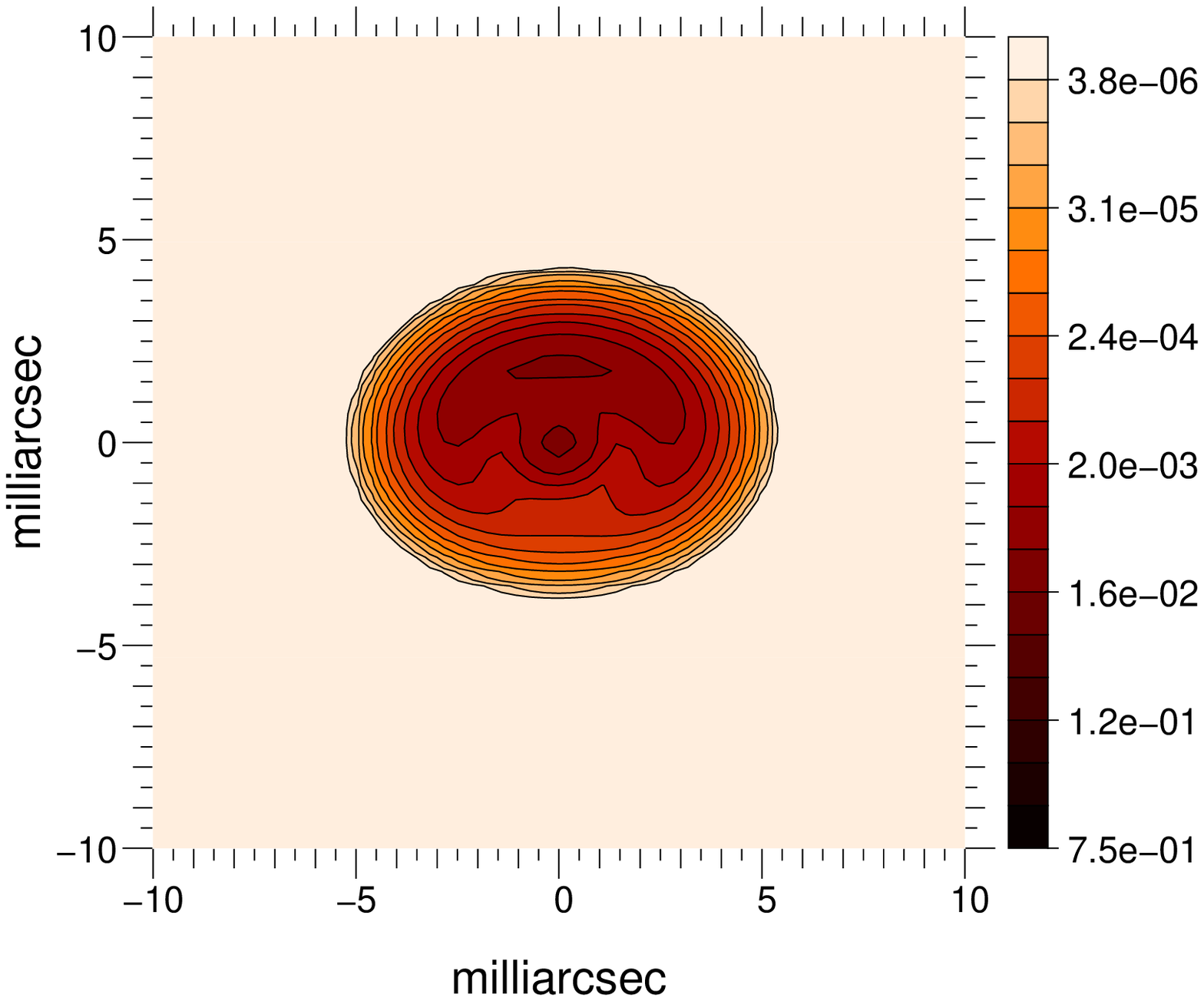}
\includegraphics[height=4cm]{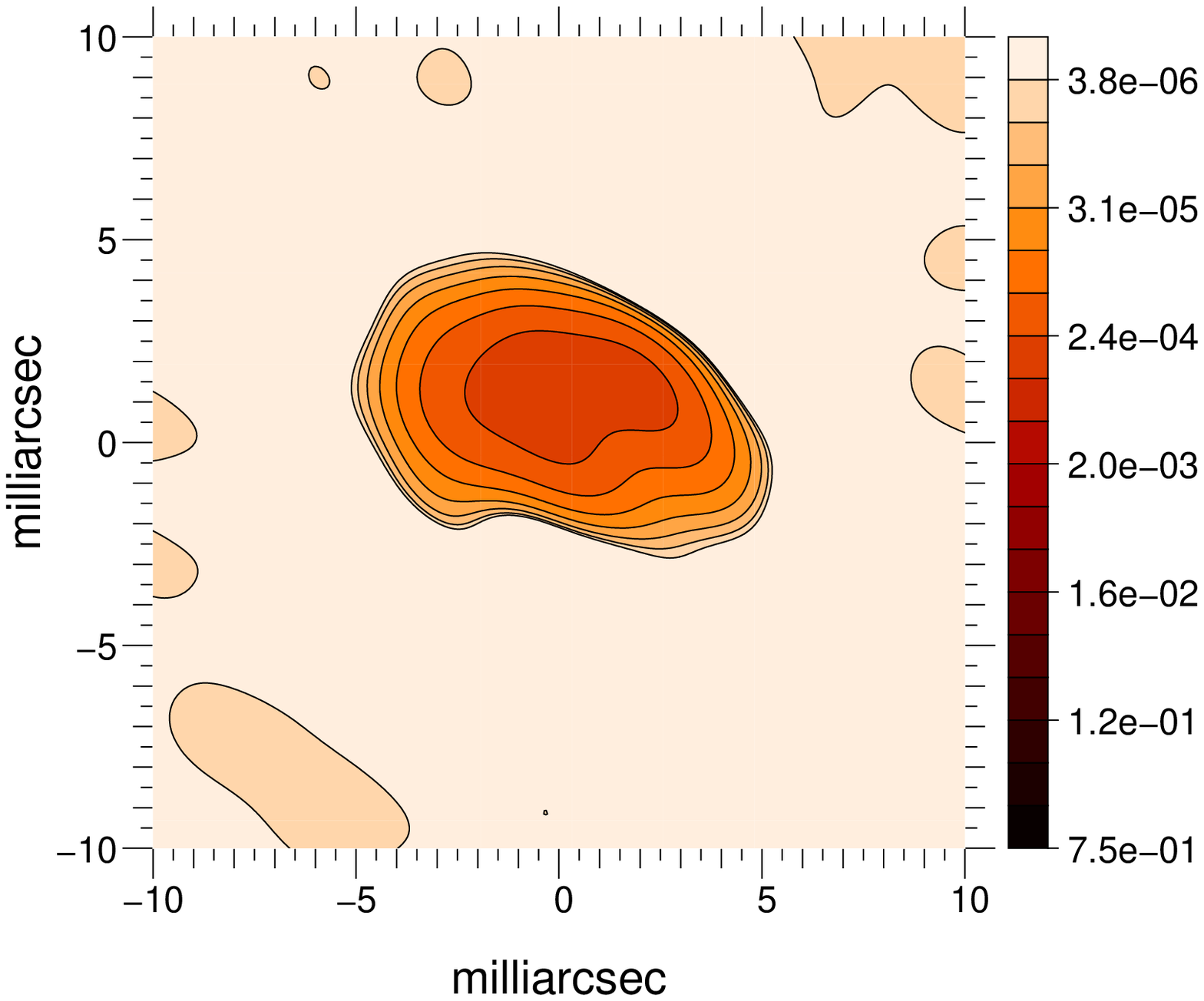}
\end{tabular}
   \end{center}
\end{figure}

\begin{figure}[h!]
   \begin{center}
   \begin{tabular}{c}
\includegraphics[height=4cm]{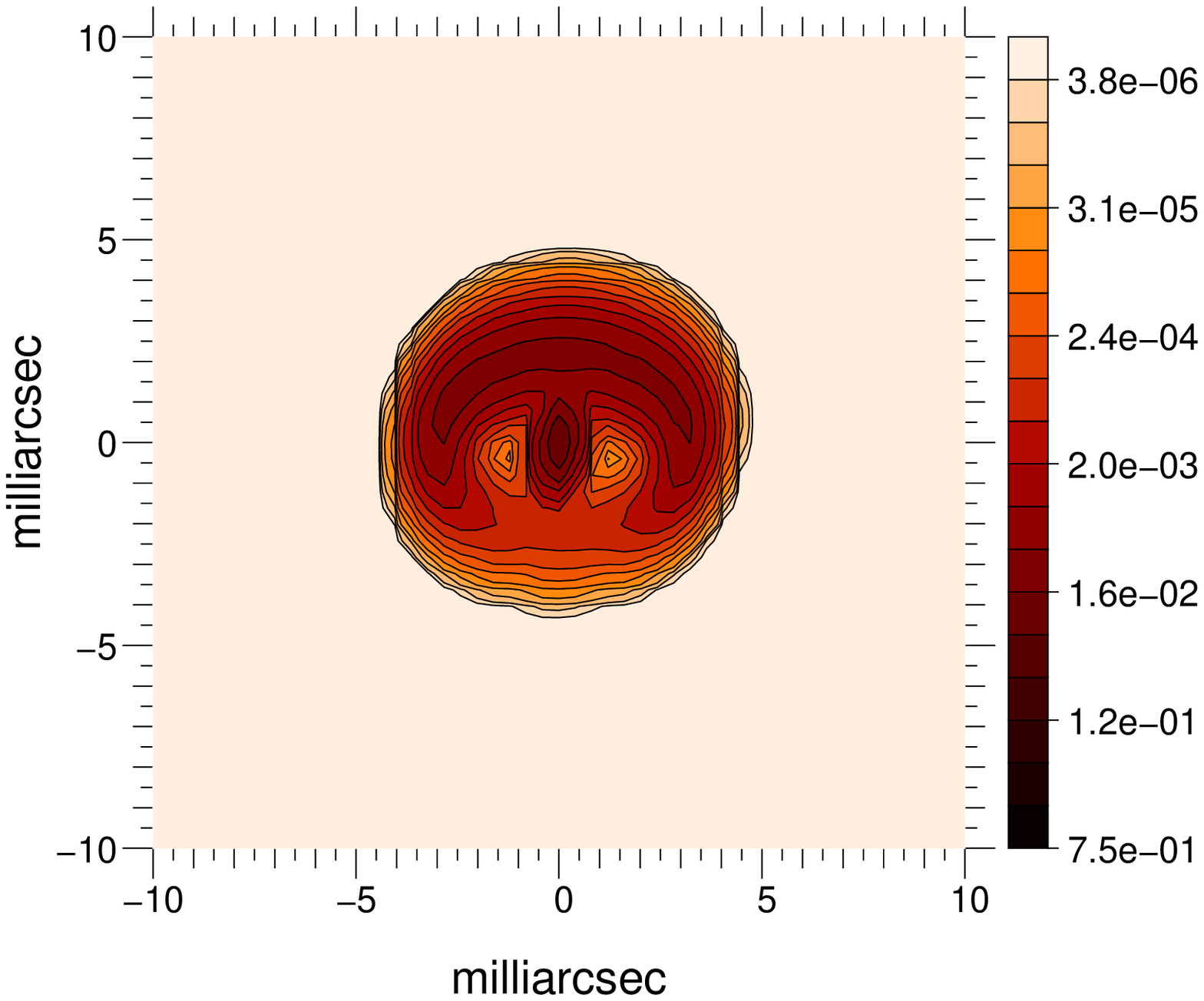}
\includegraphics[height=4cm]{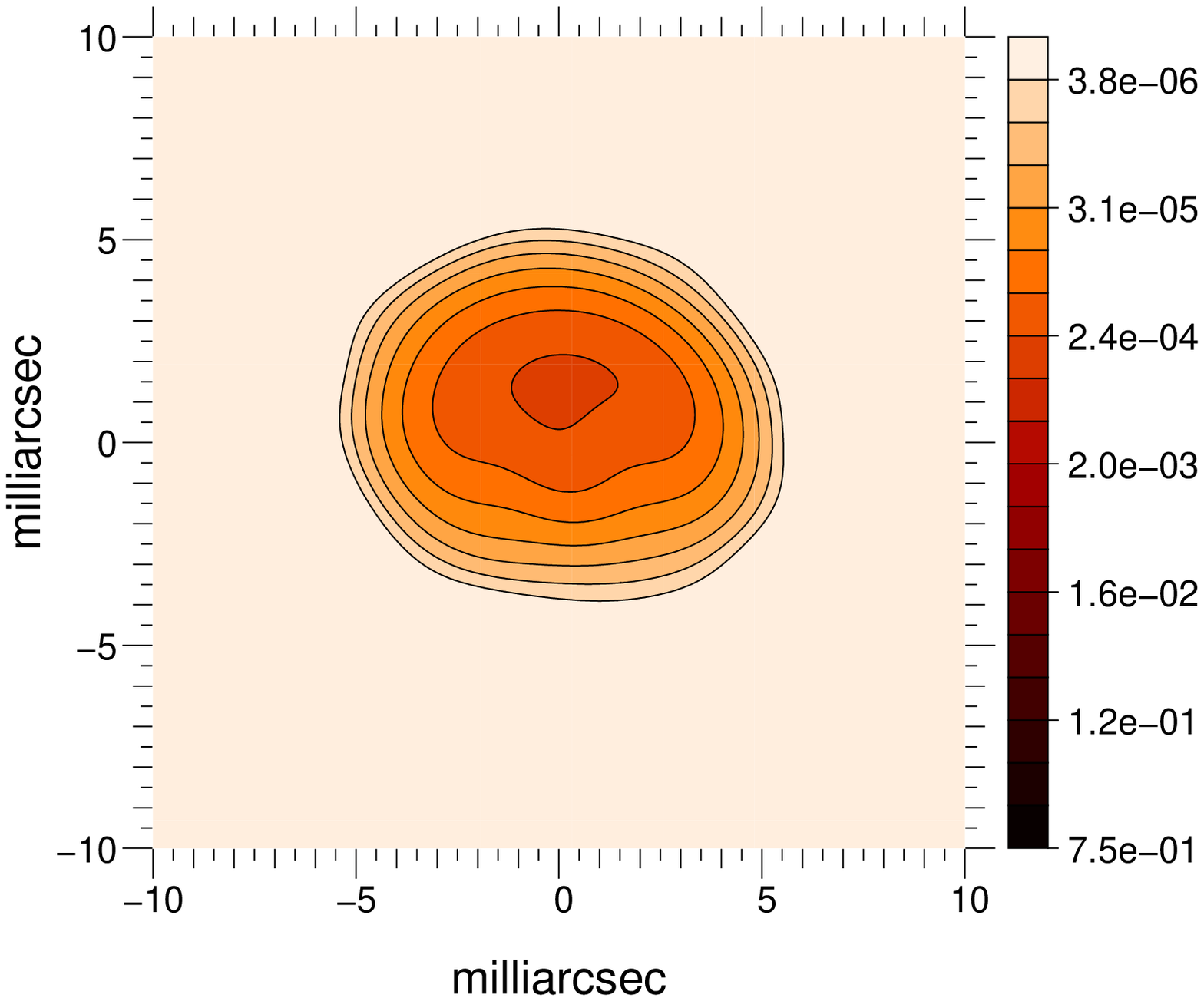}
\end{tabular}
   \end{center}
\end{figure}

\begin{figure}[h!]
   \begin{center}
   \begin{tabular}{c}
\includegraphics[height=4cm]{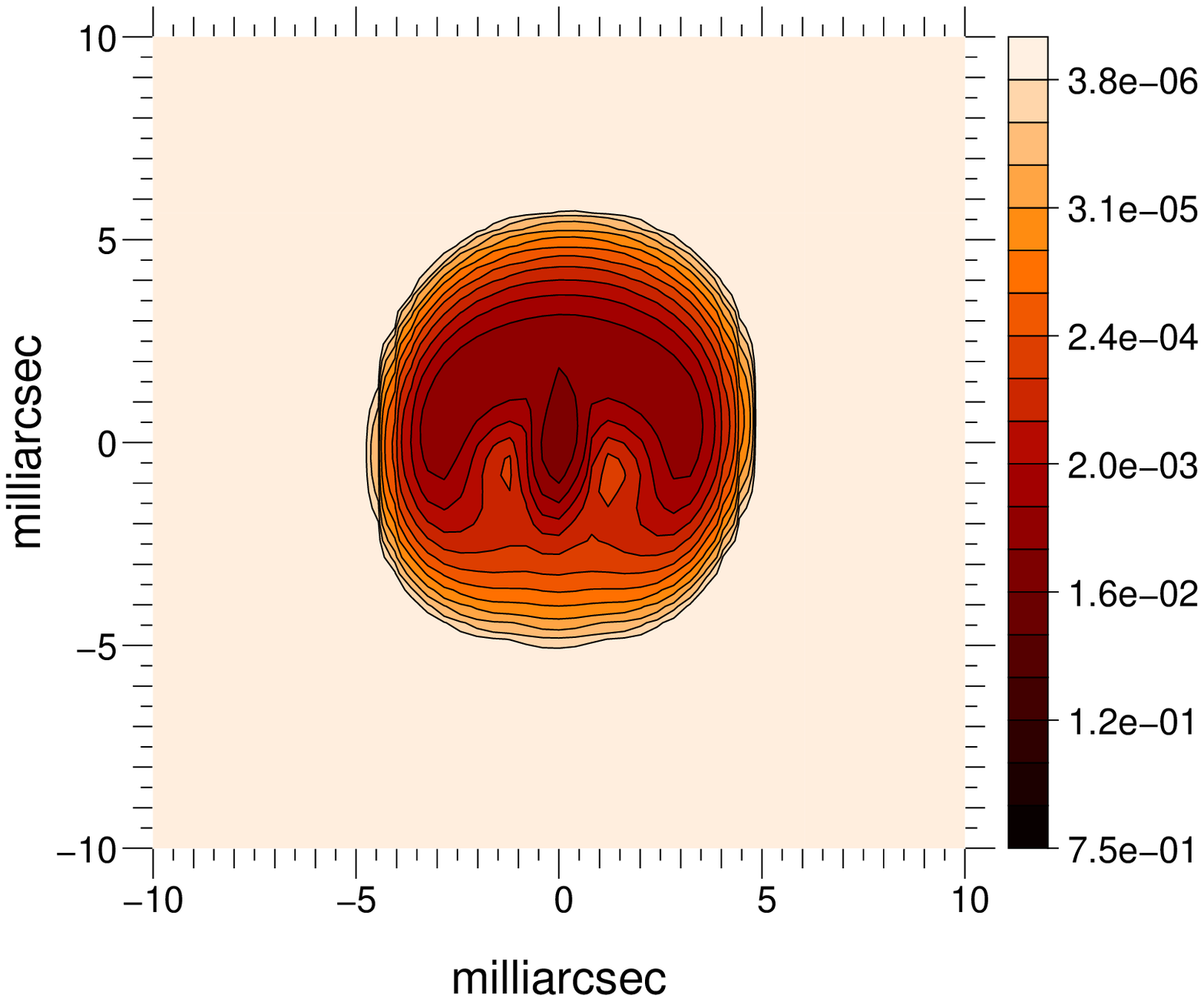}
\includegraphics[height=4cm]{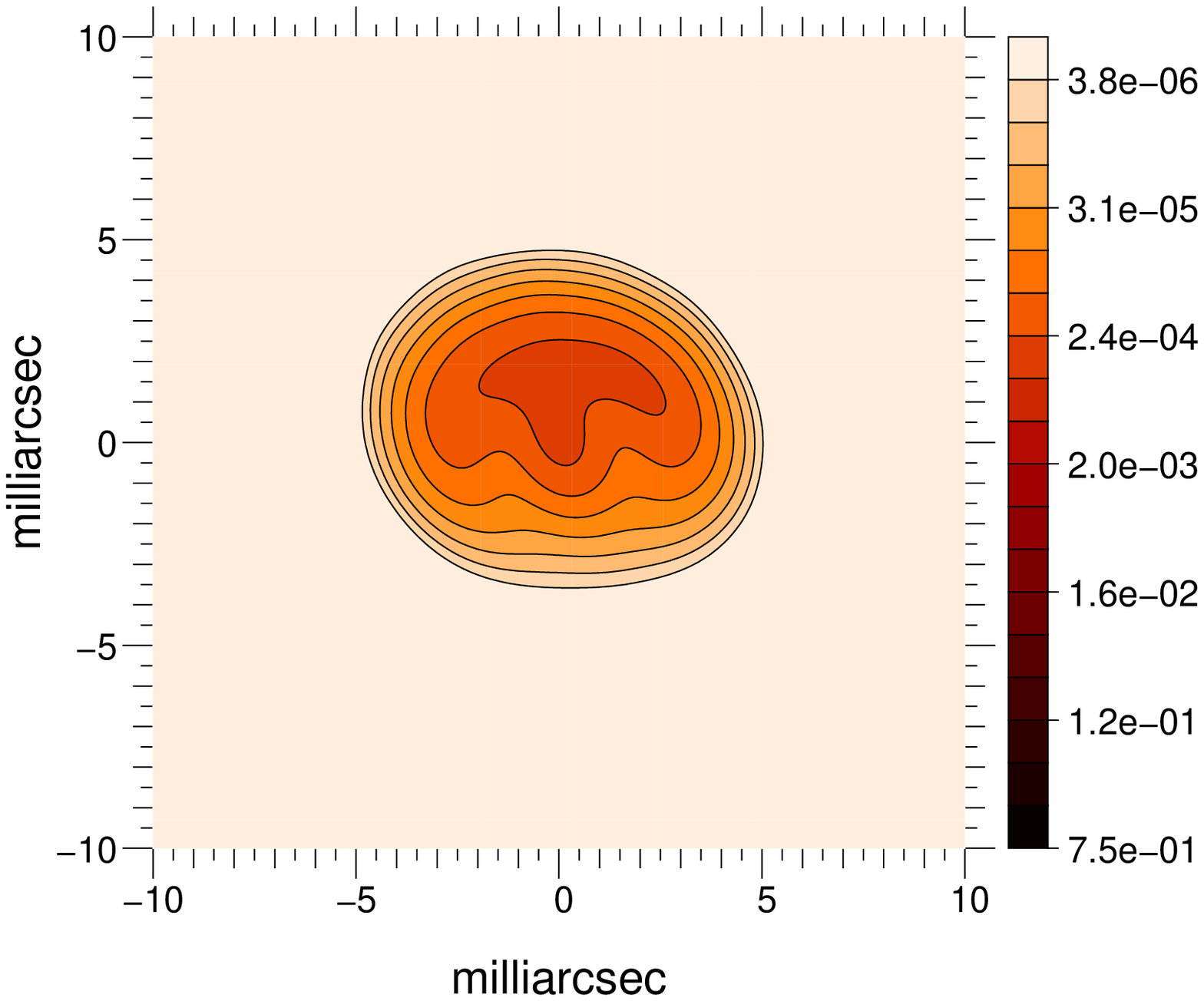}
  \end{tabular}
   \end{center}
   \caption[YSO]
   { \label{fig:example} A simulated stellar surface of the structure of inner disks surrounding YSOs, 0.1 mas/pixel sampling; convolved image, 4 UT $\times$ 1 night configuration, 0.5 mas/pixel sampling, 4.45 mas FWHM resolution; {\sc aips} reconstruction 4 UT $\times$ 1 night configuration, 0.1 mas/pixel sampling; convolved image, 4 AT $\times$ 3 nights configuration, 0.5 mas/pixel sampling, 3.51 mas FWHM resolution; {\sc aips} reconstruction 4 AT $\times$ 3 nights configuration, 0.1 mas/pixel sampling; convolved image, 6 AT $\times$ 1 night configuration, 0.5 mas/pixel sampling, 5.17 mas FWHM resolution; {\sc aips} reconstruction 6 AT $\times$ 1 night configuration, 0.1 mas/pixel sampling.}
\end{figure}

\begin{table}[h!]
\begin{center}
\footnotesize
\caption{YSO. Diameter units are pixels.}
\begin{minipage}[c]{155mm}
\begin{tabular} {l | c c c c c c}

\hline
      & Image 4 UT & {\sc aips} 4 UT & Image 4 AT 3 & {\sc aips} 4 AT 3 & Image 6 AT & {\sc aips} 6 AT \\
\hline

flux star & 15.7\% & - & 18.1\% & - & 18.1\% & 16.7\% \\

flux disk & 84.3\% & - & 81.9\% & - & 81.9\% &  83.3\% \\

ratio     & 0.2    & - & 0.2    & - & 0.2    &  0.2 \\

\hline

outer diameter & 60 $\times$ 40 & 60 $\times$ 40 & 70 $\times$ 60 & 70 $\times$ 60 & 80 $\times$ 70 & 80 $\times$ 70 \\

\hline

SNR            &  -             & 1            &      -         & 184            &   -               & 530 \\

\hline 







\end{tabular}
\end{minipage}
\end{center}
\end{table}

\section{Discussion}

If we compare the reconstructed images with the convolved images, we see that the phase referencing image reconstruction has yielded good results. Most of the flux
is recovered in the individual image elements and the astrometry is excellent. A clear example of this is the reconstructed stellar surface images. Most of the flux is recovered in a compact region and careful inspection of the reconstructed images shows correspondence with individual brightness regions in the convolved image. The edges of the stellar surface are also well reproduced.

For an array of $N$ telescopes, the total number of measurements per observing night per integration point (moduli plus phase over each baseline) in phase referencing is $N \times (N-1)$ compared to $\frac{(N-1)}{2} \times (\frac{3}{2}N-2)$ closure phases plus squared visibilities in
conventional phase closure techniques. As an example, when $N=4$, there are 6 measured parameters for phase closure observations and 12 for phase referencing. Therefore, phase referenced reconstructed images should theoretically be more rigorous.

\section{Observable Objects}

In order to perform real time phase referencing observations, it is necessary to have a phase reference
source no more than 30" away from the target source. For larger distances, the atmospheric pistons
become distinct and the calibration incorrect.

Among the science cases presented, the most
crucial in terms of number of phase reference candidates are the AGN, due to their large distances and therefore their K band faintness. We have attempted to quantify the number of AGN for which phase reference imaging is possible by using the V\'eron-Cetty \& V\'eron (2006) catalog which contains AGNs, QSOs and BL Lac objects.
Firstly, we constrained the science targets to those visible to the VLTI:
targets were chosen to be between $+$20 and $-$90 degree declination. We then searched the 2MASS point source catalog
for a nearby (less than 30" from the science target) bright star suitable for both AO wavefront sensing and fringe tracking 
(i.e., star mangitudes $K<$10, $R<$16). Science targets are listed in Table 7.


\begin{table}[h!]
\begin{center}
\footnotesize
\caption{Science target   list   of   all   southern   (declination   +20   to
-90$^{\circ}$)  AGNs in the  V\'eron-Cetty \& V\'eron catalog (2006)  with a
bright star ($K<10$ and $R<16$) in their isoplanatic patch (i.e., star
separation $<$\,30\,arcsec). These nearby  stars are suitable for both
AO wavefront  sensing and  fringe tracking. The  $K_{core}$ magnitudes
are  the $K$  magnitudes from  the 2mass  catalog, the  $K_{star}$ and
$R_{star}$ magnitudes are the magnitudes of the nearby stars.}
\begin{minipage}[c]{110mm}
\begin{tabular}{l | c c c c c c}
\hline
Name      &type & RA          & DEC    &$K_{core}$ &$K_{star}$ &$R_{star}$ \\
\hline
J004336.0+001456& Sy1 & 00 43 36.0 & +00 14 56 & 14.6 &   9.4 & 10.7 \\
LEDA101303      & LIN & 01 38 52.9 & -10 27 11 & 13.6 &   8.5 & 10.7 \\
NGC 676         & Sy2 & 01 48 57.3 & +05 54 24 & 11.1 &   8.6 & 10.1 \\
J024613.8+105656& QSO & 02 46 13.8 & +10 56 56 & 13.7 &   9.0 & 10.9 \\
NGC1204         & Sy2 & 03 04 40.0 & -12 20 29 & 11.4 &   9.1 & 10.0 \\
HE 0324-3652    & QSO & 03 26 01.0 & -36 41 49 & 13.9 &   9.7 & 12.8 \\
LEDA13424       & LIN & 03 38 40.5 & +09 58 12 & 12.4 &   8.1 & 12.6 \\
ESO548-81       & Sy1 & 03 42 03.6 & -21 14 37 & 10.5 &   6.0 &  8.4 \\
LEDA 2824014    & QSO & 04 37 36.6 & -29 54 02 & 14.3 &   7.0 &  8.0 \\
LEDA2824155     & Sy1 & 04 56 08.9 & -21 59 09 & 13.4 &   9.6 & 11.0 \\
4U0517+17       & Sy1 & 05 10 45.5 & +16 29 57 & 11.6 &   6.2 &  9.2 \\
J052223.1-072513& Sy1 & 05 22 23.1 & -07 25 13 & 13.0 &   9.5 & 10.5 \\
J062233.8-231743& Sy1 & 06 22 33.4 & -23 17 42 & 13.0 &   9.4 & 11.3 \\
J063635.8-622032& LIN & 06 36 35.8 & -62 20 32 & 14.2 &   8.4 & 10.0 \\
J083750.7+091218& Sy1 & 08 37 50.7 & +09 12 18 & 14.9 &   8.8 & 10.5 \\
2E2060          & Sy1 & 08 52 15.1 & +07 53 37 & 12.6 &   9.2 & 10.8 \\
J091034.3+031328& AGN & 09 10 34.3 & +03 13 28 & 14.0 &   9.7 & 11.3 \\
J091430.4+104906& Sy1 & 09 14 30.4 & +10 49 06 & 14.7 &   9.3 & 10.5 \\
J095916.7-073517& Sy1 & 09 59 16.7 & -07 35 17 & 13.1 &   9.2 & 11.1 \\
LEDA31718       & Sy1 & 10 39 46.3 & -05 28 59 & 12.8 &   9.2 & 11.3 \\
RBS 999         & Sy1 & 11 34 22.5 & +04 11 28 & 12.9 &   8.8 & 10.5 \\
MGC 24800       & AGN & 11 48 16.0 & -00 03 29 & 14.0 &  8.4 & 11.3 \\
J120001.9+023418& Sy2 & 12 00 01.9 & +02 34 18 & 14.5 &  9.1 & 11.2 \\
J120848.9+101343& AGN & 12 08 48.9 & +10 13 43 & 14.3 &  9.8 & 11.0 \\
J121855.8+020002& QSO & 12 18 55.8 & +02 00 02 & 14.8 &   9.6 & 11.1 \\
J130335.3-004912& Sy1 & 13 03 35.3 & -00 49 12 & 14.7 &  9.2 & 11.5 \\
HE 1304-0541    & QSO & 13 06 47.6 & -05 57 35 & 14.7 &   8.1 & 10.2 \\
J130838.2-825934& QSO & 13 08 38.2 & -82 59 34 & 14.6 &   8.7 & 11.4 \\
J132301.0+043951& BLL & 13 23 01.0 & +04 39 51 & 14.4 &   9.7 & 10.9 \\
LEDA 170317     & Sy2 & 13 58 59.7 & -20 02 43 & 12.3 &  8.2 &  8.0 \\
J152929.3+033137& Sy2 & 15 29 29.3 & +03 31 37 & 14.7 &  9.1 & 11.2 \\
MCG+03-40-009   & Sy2 & 15 35 52.6 & +14 31 04 & 12.9 &   9.6 & 12.5 \\
J154025.1+030640& Sy2 & 15 40 25.1 & +03 06 40 & 15.0 &  9.4 & 11.3 \\
ESO 137-34      & Sy2 & 16 35 14.2 & -58 04 41 & 11.4 &   7.3 &  9.2 \\
LEDA 2829294    & QSO & 17 33 02.6 & -13 04 50 & 14.2 &   7.4 & 14.8 \\
J173728.3-290802& Sy1 & 17 37 28.3 & -29 08 02 & 11.1 &  8.7 & 11.0 \\
IGR J18027-1455 & Sy1 & 18 02 47.3 & -14 54 54 & 10.9 &   8.6 & 15.2 \\
LEDA 86291      & Sy1 & 18 51 59.5 & +11 52 33 & 11.0 &   9.9 & 14.2 \\
J193109.5+093713& BLL & 19 31 09.8 & +09 37 04 & 14.4 &   8.9 & 13.5 \\
LEDA 65714      & Sy1 & 20 55 22.3 & +02 21 17 & 12.5 &   9.6 & 12.7 \\
1H 2107-097     & Sy1 & 21 09 09.9 & -09 40 15 & 10.9 &   8.8 & 12.1 \\
J211837.3-010537& AGN & 21 18 37.3 & -01 05 37 & 8.14 &   8.1 & 12.3 \\
J220555.0-000755& Sy2 & 22 05 55.0 & -00 07 55 & 14.8 &   8.9 & 11.6 \\
J223013.4-292554& QSO & 22 30 13.4 & -29 25 54 & 14.9 &   9.8 & 11.0 \\
MCG+01-57-007   & Sy1 & 22 32 30.8 & +08 12 27 & 11.8 &   9.3 & 10.7 \\
\hline
\end{tabular}
\end{minipage}
\end{center}
\end{table}



We have also investigated the number of phase reference targets available from
a series of stellar catalogs. No trimming of objects for
observability (southern hemisphere or target magnitude) was done.
The following catalogs were searched: the de~Winter et al.~(2001)
catalog of southern emission line objects mainly containing 
Herbig Ae/Be stars and some B[e], LBV and TTauri stars; the
Egret~(1980) catalog of supergiants; the Fracassini et al.~(1994)
catalog of stellar radii trimmed to objects with diameters of at
least 10~mas; the Ramos-Larios \& Phillips~(2005) catalog of
planetary nebulae; the Herbig \& Bell~(1995) catalog of
pre-main-sequence stars; the Muench et al.~(2002) catalog of
pre-main-sequence stars in the Orion Trapezium cluster. 
We have searched for appropriate phase reference 2MASS point sources 
with K band magnitudes between 9 and 11 less then 30'' away from the
targets. The results
are presented in Table~8 and highlight the potencial of phase
reference imaging.

\begin{table}[h!]
\begin{center}
\footnotesize \caption{Science target   list  with a bright star (9$<$K$<$11)
in their isoplanatic patch (i.e., star
separation $<$\,30\,arcsec). These nearby  stars are suitable for 
fringe tracking. }
\begin{minipage}[c]{170mm}
\begin{tabular} {l | c | c | c}

\hline

Catalog  & Science & Total Number of Targets & Total Number of Targets \\
Reference & Targets & in the Catalog          & with Phase Ref. Sources \\

\hline

de Winter et al.& Herbig Ae/Be, LBV \& T Tauri &  162  &62\\
                &                               &       & \\
Egret &supergiants & 5073   &1253\\
      &            &        &    \\
Fracassini et al.& stars with d$<$10 mas & 143 &8\\
                 &                       &     & \\
Ramos-Larios \& Phillips & planetary nebulae &325 &68\\
                         &                   &    &  \\
Herbig \& Bell & pre-main sequence stars & 763  &217\\
               &                         &      & \\
Muench et al.& pre-main sequence stars &          &  \\
             & in the Orion Trapezium cluster & 1 010  & 875\\
             &                                &       &   \\
V\'eron-Cetty \& V\'eron & AGN, BL Lacs \& QSOs & 108 080 &  133 \\
\hline

\end{tabular}
\end{minipage}
\end{center}
\end{table}





\section {Conclusions}

Theoretically it is found that phase reference image reconstruction should yield more rigourous images than
the classical phase closure method used in optical interferometry, not only because longer integrations
times are allowed, but also because the former method measures information related to the visibility phase.

We wished to explore the potential of phase referencing in optical interferometry. We have generated simulated, noise visbility amplitudes and phases using the VLTI as a template and reconstructed images with the classic radio interferometry algorithm {\sc clean} contained in the {\sc aips} package.
Our results show that with this method we will be able to reconstruct images of diverse sources with
a spatial resolution of about 4 milliarcseconds in K band.

We have also compiled a list of target sources for which optical phase referencing is possible and found several hundred 
candidates.

\section{Acknowledgments}

We would like to thank J. Brinchmann for his help with the phase
reference sources. MEF is supported by the Funda\,c\~ao para a Ci\^encia e a Tecnologia through the research grant
SFRH/BPD/36141/2007. PJVG and MEF were supported in part by the
Funda\,c\~ao para a Ci\^encia e a Tecnologia through projects
PTDC/CTE-AST/68915/2006 and PTDC/CTE-AST/65971/2006 from POCI, with
funds from the European programme FEDER.

\section {References}






\noindent Egret, D.\ 1980, Bulletin d'Information du Centre de
Donnees Stellaires, 18, 82

\noindent Fracassini, M., Pasinetti-Fracassini, L.~E., Pastori, L.,
\& Pironi, R.\ 1994, VizieR Online Data Catalog, 2155, 0

\noindent Garcia, P. et al. 2007, in doc. VLT-SPE-VSI-15870-4335, issue 1.0 in VSI Phase A Document Package, Science Cases

\noindent Herbig, G.~H., \& Bell, K.~R.\ 1995, VizieR Online Data
Catalog, 5073, 0

\noindent H\"{o}gbom, J. A., 1974, A\&AS, 15, 417

\noindent Jocou, L. et al. 2007, in doc. VLT-SPE-VSI-15870-4335, issue 1.0 in VSI Phase A Document Package, System Design


\noindent Masoni, L. 2006, PADEU, 17, 155

\noindent Masoni, L. et al. 2005, Astr. Nachr., 326, 566

\noindent Muench, A.~A., Lada, E.~A., Lada, C.~J., \& Alves, J.\
2002, ApJ, 573, 366

\noindent Ramos-Larios, G., \& Phillips, J.~P.\ 2005, MNRAS, 357,
732



\noindent V\'eron-Cetty, M. -P. \& V\'eron, P. 2006, A\&A, 455, 773

\noindent de Winter, D., van den Ancker, M.~E., Maira, A., Th{\'e},
P.~S., Djie, H.~R.~E.~T.~A., Redondo, I., Eiroa, C., \& Molster,
F.~J.\ 2001, A\&A, 380, 609

\end{document}